%% file: IFB.tex
\documentclass[12pt,journal,draftcls,onecolumn]{IEEEtran}
\usepackage{amsmath,amssymb,mathrsfs}
\usepackage{epsfig,epsf,subfigure,graphicx,graphics}
\usepackage{url}
\usepackage{color}

\newtheorem{theorem}{Theorem}[section]

\newtheorem{remark}{Remark}[section]
\newtheorem{claim}{Claim}[section]

\newcommand{\lp}{\left(}
\newcommand{\rp}{\right)}

\newcommand{\lbp}{\left\{}
\newcommand{\rbp}{\right\}}

\newcommand{\msf}{\mathsf}

\graphicspath{{../figs/}}

\IEEEoverridecommandlockouts

\allowdisplaybreaks

\usepackage{cite,subfigure}
\usepackage{algorithmic,algorithm}
\usepackage{xcolor}
\usepackage{comment}

\newcommand{\dl}{\delta}

\title{Erasure Broadcast Channels with\\ Intermittent Feedback}

\author{Alireza~Vahid, 
~Shih-Chun~Lin,
~I-Hsiang~Wang
\thanks{Alireza Vahid is with the Department of Electrical Engineering, University of Colorado Denver, Denver, CO, USA. Email: {\sffamily alireza.vahid@ucdenver.edu}. Shih-Chun Lin is with Department of Electrical and Computer Engineering, NTUST, Taipei, Taiwan. Email: {\sffamily sclin@ntust.edu.tw}. I-Hsiang Wang is with Department of Electrical Engineering, National Taiwan University, Taipei, Taiwan. Email: {\sffamily ihwang@ntu.edu.tw}.}
\thanks{Preliminary results of this work were presented at the 2019 IEEE Information Theory Workshop (ITW)~\cite{ITW-IFB}.}
}

\begin{document}
\maketitle


\begin{abstract}
Achievable data rates in wireless systems rely heavily on the available channel state information (CSI) throughout the network. However, feedback links, which provide this information, are scarce, unreliable, and subject to security threats. In this work, we study the impact of having intermittent feedback links on the capacity region of the canonical two-user erasure broadcast channels. In our model, at any time instant, each receiver broadcasts its CSI, and at any other node, this information either becomes available with unit delay or gets erased. For this setting, we develop a new set of outer bounds to capture the intermittent nature of the feedback links. These outer bounds depend on the probability that the CSI from both receivers are erased at the transmitter. In particular, if at any time, the CSI from at least one of the two receivers is available at the other two nodes, then the  outer-bounds match the capacity with global delayed CSI. We also provide capacity-achieving transmission strategies under certain scenarios, and we establish a connection between this problem and Blind Index Coding with feedback. 
\end{abstract}

\begin{IEEEkeywords}
Erasure broadcast channel, intermittent feedback, Shannon feedback, capacity region, delayed CSI.
\end{IEEEkeywords}


\section{Introduction}
\label{Section:Introduction_IFB}

\input{Introduction_IFB.tex}


\section{Problem Formulation}
\label{Section:Problem_IFB}

\input{Problem_IFB.tex}


\section{Statement of the Main Results}
\label{Section:Main_IFB}

\input{Main_IFB_symmetric.tex}


\section{Proof of Theorem~\ref{THM:Capacity_IFB}: Outer Bounds}
\label{Section:Converse_IFB}

\input{Converse_IFB.tex}


\section{Proof of Theorem~\ref{THM:Ach_IFB}: Transmission Protocol}
\label{Section:Achievability_IFB}

\input{Achievability_IFB_March2020.tex}

\section{Proof of Theorem~\ref{THM:InnerBound_IFB}}
\label{Section:Achievability_InnerBound_IFB}

\input{Achievability_InnerBound_IFB.tex}


\section{Discussion on Other Regimes and Connection to Blind Index Coding}
\label{Section:Discussion_IFB}

The authors conjecture that the outer-bound region of Theorem~\ref{THM:Capacity_IFB} is also fact the capacity region of the two-user erasure broadcast channel with intermittent feedback considered in Theorem \ref{THM:InnerBound_IFB}. We believe the transmission strategy needs to be improved to provide a smooth transition between global delayed CSI scenario and the one-sided scenario. As we discussed briefly in Section~\ref{Section:Achievability_IFB}, the transmission strategy under the one-sided feedback assumption~\cite{ISIT19sclin} requires careful utilization of the available feedback in all phases of communication. However, in Case 2 of Theorem~\ref{THM:Ach_IFB} and in the proof of Theorem~\ref{THM:InnerBound_IFB}, we do not utilize feedback during certain phases of communications. The reason why the achievable rates do not degrade by ignoring feedback during Phase 3 of each iteration of Case~2 lies in the specific assumptions on channel parameters: we assume fully correlated feedback links and thus, transmitter wither has delayed feedback from both receivers or from no one.

The story is more complicated for Theorem~\ref{THM:InnerBound_IFB}. In the BIC phase, we have a BC in which a fraction of the bits for each receiver is available to the other one but the transmitter does not know which ones. However, different to the BIC problem in \cite{kao2016blind}, now we have additional (intermittent) feedback. To simplify the scheme, we provide signals intended for each receiver to both. However, using feedback one can indeed improve upon this scheme and improve the rates. Unfortunately, we face the following challenges: (1) the scheme becomes rather complex even if we assume after the BIC step feedback links are no longer intermittent; (2) even for erasure BIC without feedback, the capacity is not provided in \cite{kao2016blind}. Not to mention our more complicated setting. Thus, the BIC with feedback is an interesting yet challenging problem that needs to be solved before we can fully address the capacity region of the two-user erasure broadcast channel with intermittent feedback. Ideally, the transmission strategy should cover both extreme cases of DD and DN, and this is an ongoing part of our research.


\section{Conclusion}
\label{Section:Conclusion_IFB}

We developed new outer bounds on the capacity region of two-user erasure BCs with intermittent feedback. We also showed that these outer bounds are achievable under certain assumptions on channel parameters. The next step is to characterize the capacity region for all channel parameters. We conjecture the outer bounds are tight, and the region given in Theorem~\ref{THM:Capacity_IFB} is in fact the capacity region. However, we need to solve BIC with (intermittent) feedback as an intermediate step before settling this conjecture. An interesting future direction is consider the scenario in which receivers can encode the feedback messages, and find the minimum required feedback bandwidth to achieve the global feedback performance with only intermittent feedback.

Finally, in~\cite{vahid2016two}, two-user erasure interference channels~\cite{vahid2014capacity} with local delayed feedback were studied. There, it was shown that each transmitter must at least know the delayed CSI of the links connected to it in order to achieve the global delayed CSI performance. For distributed transmitters, understanding the capacity region of the two-user erasure interference channel with intermittent feedback would be the starting point.

\bibliographystyle{ieeetr}
\bibliography{bib_IFB}

\end{document}

%% file: Introduction_IFB.tex
{Data rates, reliability, and stability of wireless systems rely heavily on the availability of channel state information (CSI) throughout the network. This information could be as simple as ACK/NACK packets sent back via the feedback links. However, there are challenges in constantly providing CSI to all nodes. For instance, in large-scale systems and massive multiple-input multiple-output (MIMO) networks, feedback locality is the reality~\cite{de2012degrees,vahid2017interference,deng2019intermittent}. One could also use a side-channel such as best-effort WiFi for feedback in which CSI packets may sometimes get dropped~\cite{karakus2015}. Moreover, feedback channels are typically low bandwidth and thus, unprotected against security threats, which has enabled a class of security attacks to target the CSI signals in order to disrupt communications~\cite{sadeghi2018adversarial,liu2020adversarial}. Thus in this paper, we will investigate how to exploit intermittent CSI feedback to improve network capacity.}



{In this work, we consider the classical two-user erasure broadcast channels (BCs) with intermittent channel state feedback to provide a fundamental understanding of communications in such scenarios. In a packet-based communication network, each hop can be modeled as a packet erasure channel~\cite{DanaGowaikar_06}, and thus, studying the erasure BCs provides a good understanding of multi-session uni-casting in small wireless networks~\cite{GeorgiadisDetDelayedBC,Wang_12,GatzianasGeorgiadis_13}.} We consider an erasure BC with a transmitter and two  receivers.
The channel model from the transmitter to the two receivers follows the standard erasure BC model. For the feedback model, at any time instant, each receiver $i$, $i=1,2$, broadcasts its CSI to the other two nodes, that is, whether or not the symbol sent by the transmitter successfully arrives. Then, with some probability, this broadcast of CSI is successful at each of the other nodes, and the transmitter and/or the other receiver will learn the CSI of receiver $i$ with unit delay; otherwise, the feedback signal is erased. We refer to this model as the two-user erasure BC with {\it intermittent feedback}. This model generalizes those in prior works which either assume all receivers can provide delayed state feedback~\cite{GeorgiadisDetDelayedBC,Wang_12,GatzianasGeorgiadis_13}, only a single user provides its CSI~\cite{sc2016ISIT,HeYang2017ITW,sclin2018IT,ISIT19sclin}, or assume partial feedback is available~\cite{dueck1980partial,shayevitz2012capacity,venkataramanan2013achievable}. In the context of erasure BCs, in~\cite{ISIT19sclin}, we showed that the capacity region of the two-user erasure BC with global delayed CSI can be achieved with single-user delayed CSI only. This is result is in sharp contrast to the continuous channel model where (asymptotic) capacity collapses to that of no CSI when a single feedback link is missing.

{The initial steps if this work were taken in~\cite{ITW-IFB} where a more restrictive model for intermittent feedback where the delayed CSI of each receiver would either be available to {\it all} other nodes with delay or it would get erased at {\it all} other nodes.} This could correspond to a faulty feedback mechanism at the receiver where the signal either goes out or gets dropped. We improve upon the results of~\cite{ITW-IFB}, both in terms of the outer-bounds and the achievable region. More precisely, for the outer-bounds, we allow for the erasure feedback links initiating at each receiver to have a general distribution. Moreover, we provide a broader set of conditions under which these outer-bounds can be achieved, thus, characterizing the capacity region in those cases.

Our contributions in this work are two-fold. First, we derive a new set of outer bounds the two-user erasure BC with intermittent feedback. The derivation has two stages. In the first stage, we create a modified BC in which forward links are fully correlated across users when the CSI from each receiver is erased at least one other node. We show that the capacity region of this modified problem is the same as that of the original problem. In the second stage, we derive the outer bounds for this modified problem using an extremal entropy inequality for erasure links with delayed CSI. These outer bounds are governed by the probability of missing at least one feedback link from each receiver. In other words, as long as one receiver's CSI is avaialble at any given time, the outer-bound region does not degrade compared to the one with global delayed CSI. This observation matches our earlier findings in~\cite{ISIT19sclin} where the capacity region of two-user erasure BC with global delayed CSI could be achieved with single-user delayed CSI.

We also show that under certain conditions these outer bounds can be achieved. One such scenario is when the feedback links from the two receivers are fully correlated, that is, they are either both available or both erased. We propose a recursive transmission strategy where the first iteration has three phases resembling the three phase communication protocol of the global feedback case~\cite{GeorgiadisDetDelayedBC}. After these three phases, we create recycled bits that we feed to the same transmission protocol as the recursive step.  When all forward and feedback links have the same erasure probability, we provide an achievable rate region that comes close to the outer-bounds. However, as we discuss in Sections~\ref{Section:Achievability_InnerBound_IFB} and~\ref{Section:Discussion_IFB}, improving the rates requires solving Blind Index Coding (BIC) with (intermittent) feedback. The BIC problem was introduced in~\cite{kao2016blind} with any feedback, but even in that case, the problem is complicated and the capacity remains unknown.

It is worth comparing our results to another line of work in which the availability of CSI alternates between various states~\cite{tandon2013synergistic,mukherjee2017secure,chen2014vector}. However, these results assume at each time the CSI availability structure is known to the transmitter, whereas in our work, the transmitter does not know whether or not the CSI of the current transmission will be available since the feedback links are erased randomly at each time.

The rest of the paper is organized as follows. We describe the problem formulation in Section~\ref{Section:Problem_IFB}. We state our main contributions in Section~\ref{Section:Main_IFB}. The proofs are presented in Sections~\ref{Section:Converse_IFB} and~\ref{Section:Achievability_IFB}. Section~\ref{Section:Conclusion_IFB} concludes the paper. 

%% file: Problem_IFB.tex
\begin{figure}[!ht]
\centering
\subfigure[]{\includegraphics[width = .4\columnwidth]{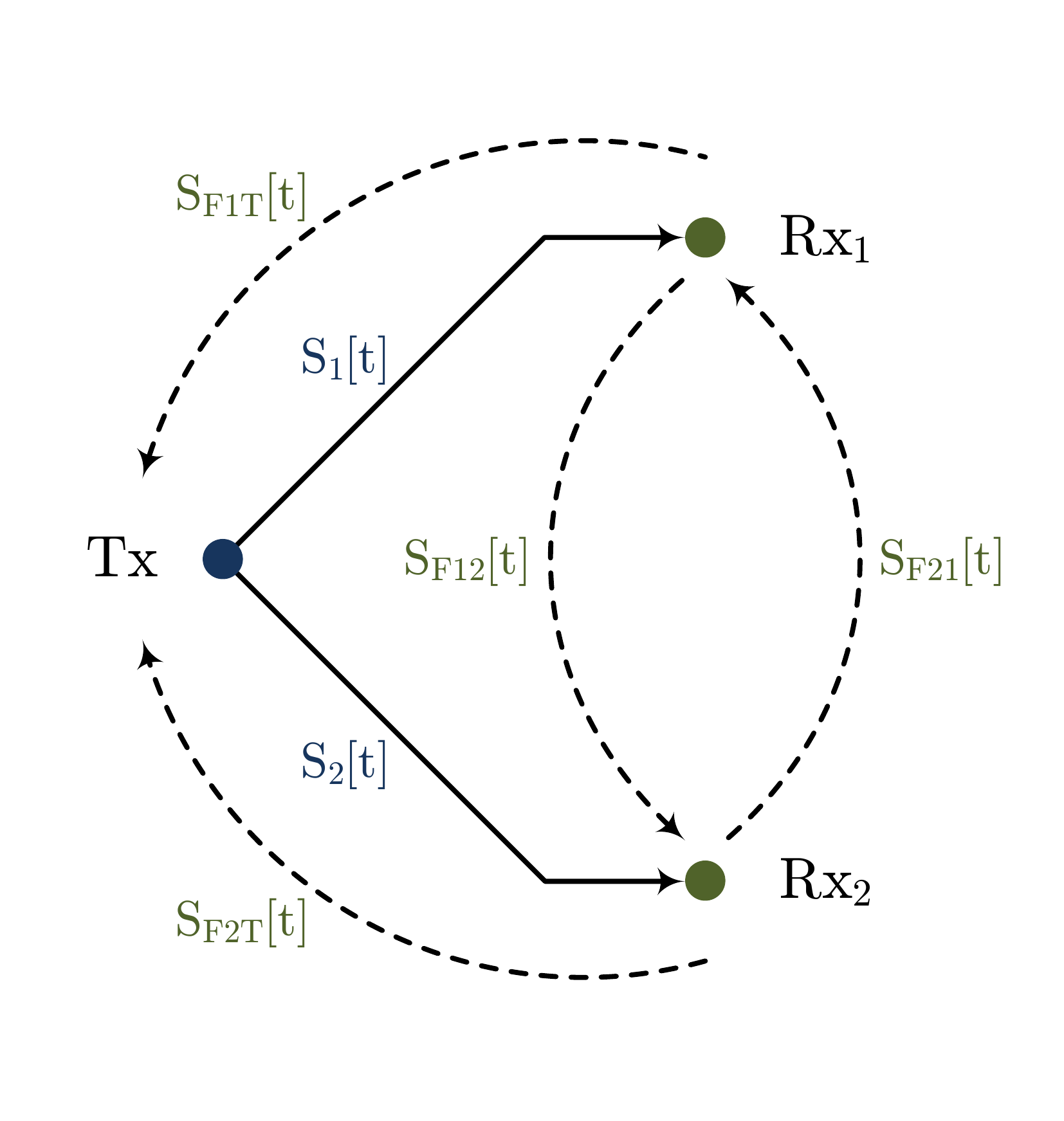}}
\hspace{.25in}
\subfigure[]{\includegraphics[width = .4\columnwidth]{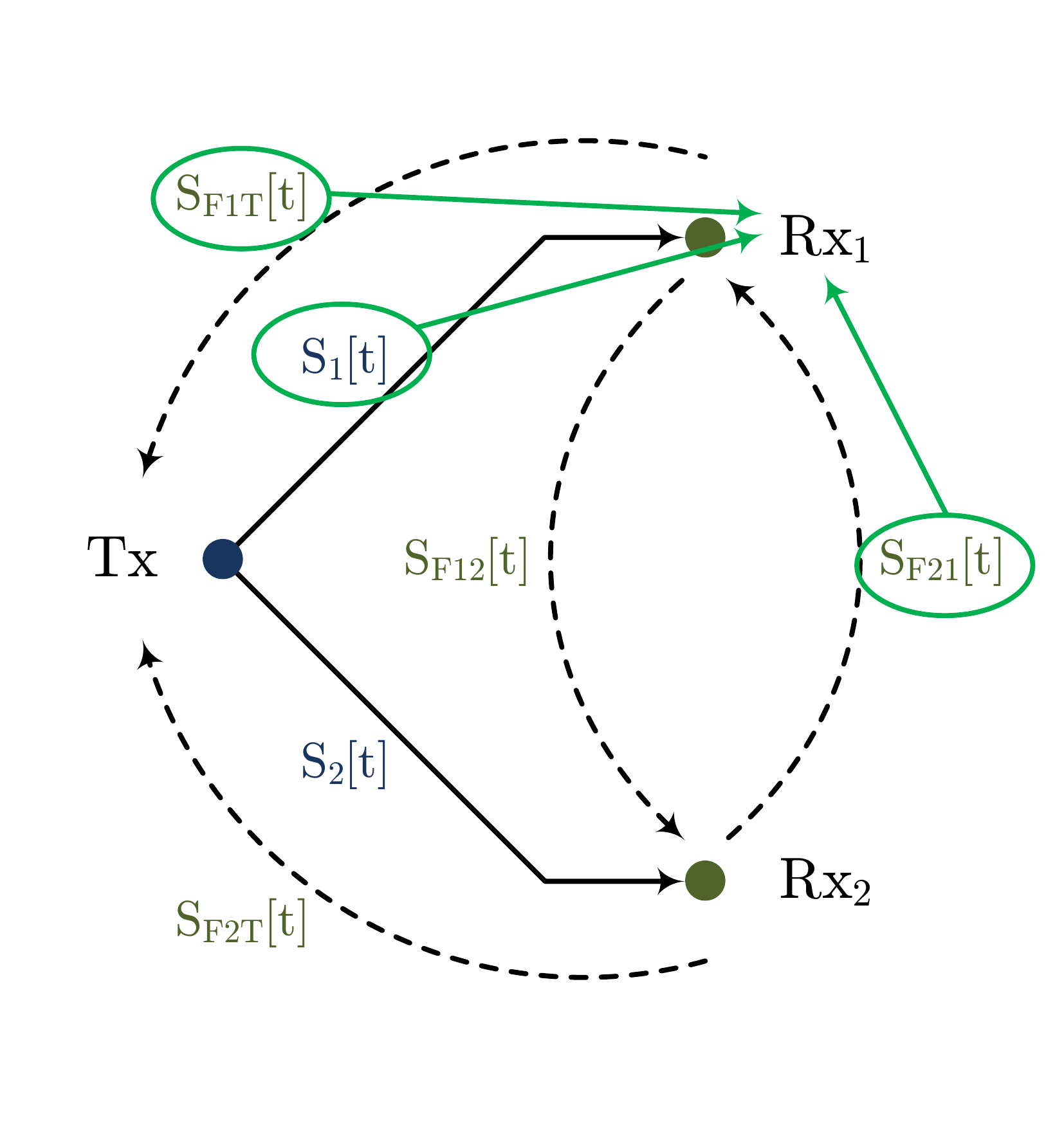}}
\caption{(a) Two-user erasure broadcast channel with intermittent feedback; (b) Available CSI at $\msf{Rx}_1$ as discussed in Remark~\ref{Remark:CSIR}. \label{Fig:BC_IFB}}
\end{figure}

We consider the two-user erasure broadcast channel of Fig.~\ref{Fig:BC_IFB}(a) in which a single-antenna transmitter, $\msf{Tx}$, wishes to transmit two independent messages, $W_1$ and $W_2$, to two single-antenna receiving terminals $\msf{Rx}_1$ and $\msf{Rx}_2$, respectively, over $n$ channel uses. Each message, $W_i$, is uniformly distributed over $\lbp1,2,\ldots,2^{nR_i}\rbp$, for $i=1,2$. At time instant $t$, the messages are mapped to channel input $X[t] \in \mathbb{F}_2$, the binary field, and the corresponding received signals at $\msf{Rx}_1$ and $\msf{Rx}_2$ are
\begin{align}
\label{eq_DL_channel}
Y_1[t] = S_1[t] X[t]~~ \; \mbox{and} \;~~ Y_2[t] = S_2[t] X[t],
\end{align}
respectively, where $\lbp S_i[t]\rbp$ denotes the Bernoulli $(1-\delta_i)$ process that governs the erasure at $\mathsf{Rx}_i$, and it is distributed i.i.d. over time.

We assume that each receiver is aware of its CSI at time $t$. When $S_i[t]=1$, $\mathsf{Rx}_i$ receives $X[t]$ noiselessly; when $S_i[t]=0$, it receives an erasure. More precisely, since receiver $\msf{Rx}_i$ knows the value of $S_i[t]$, it can map the received signal for $S_i[t] = 0$ to an erasure. While forward channels are distributed independently over time, we assume a general distribution across users, and in particular, we assume
\begin{align}
\label{Eq:DeltaFF}
\delta_{12} \overset{\triangle}= P\{S_1[t]=0,S_2[t]=0\}.
\end{align}
From basic probability, we have
\begin{align}
\label{Eq:d12condition}
\max \{ \delta_1 + \delta_2 -1, 0 \} \leq \delta_{12} \leq \min \{ \delta_1, \delta_2 \}.
\end{align}

We further assume that at time instant $t$, $\mathsf{Rx}_i$ broadcasts its channel state as depicted in Fig.~\ref{Fig:BC_IFB}(a), \emph{i.e.} $S_i[t]$, and the successful delivery of this information is governed by the Bernoulli $(1-\delta_{FiT})$ process, $\lbp S_{FiT}[t]\rbp$, to the transmitter and the Bernoulli $(1-\delta_{Fi\bar{i}})$ process, $\lbp S_{Fi\bar{i}}[t]\rbp$, to the other receiver, $i=1,2$. \footnote{Similar to~\cite{karakus2015}, we assume the receivers are passive and they simply feed back CSI without any processing through a unit-delay erasure channel. This model is motivated by the fact that that receivers do not have a-priori information about the time-varying feedback channel statistics to perform active coding. Moreover, active coding will result in longer delays. If active coding is used, the intermittent feedback model reduces to the rate-limited model of~\cite{vahid2011interference}, which is beyond the scope of this paper.} These processes are distributed i.i.d. over time, and are independent of the forward channel processes. We assume a general joint distribution for $\lbp S_{F1T}[t], S_{F12}[t], S_{F2T}[t], S_{F21}[t] \rbp$, and we define
\begin{align}
\label{Eq:deltadefinitions}
&\delta_{F1} \overset{\triangle}= P\{ S_{F1T}[t]S_{F12}[t] =0 \}, \nonumber \\
&\delta_{F2} \overset{\triangle}= P\{ S_{F2T}[t]S_{F22}[t] =0 \}, \nonumber \\
&\delta_{FF} \overset{\triangle}= P\{ S_{F1T}[t] = S_{F2T}[t] =0 \}.
\end{align}
In essence, $\delta_{Fi}$ is the probability that at any time instant at least one outgoing feedback link from ${\sf Rx}_i$ is erased. Then, $\delta_{FF}$ is the probability that at any time instant both feedback links to the transmitter are erased.


\begin{remark} [Comparison to~\cite{ITW-IFB}] \label{Remark:FBDistribution}
In our initial results for this work~\cite{ITW-IFB}, the feedback links from each receiver were on or off together, meaning that the delayed CSI of each receiver would either be available to {\it all} other nodes with delay or it would get erased at {\it all} other nodes. 
The model assumed in this work generalizes that of~\cite{ITW-IFB} to the scenario in which the erasure feedback links initiating at each receiver to have a more general distribution.
\end{remark}

The constraint imposed at the encoding function $f_t(.) $ at time index $t$ is
\begin{align}
\label{eq_enc_function}
X[t] = f_t\lp W_1, W_2, S_{F1T}^{t-1}, S_{F2T}^{t-1}, \lbp S_{F1T}S_1 \rbp^{t-1}, \lbp S_{F2T}S_2 \rbp^{t-1} \rp,
\end{align}
where
\begin{align}
S_{FiT}^{t-1} &= (S_{FiT}[1],\ldots,S_{FiT}[t-1]), \\
\lbp S_{FiT}S_i \rbp^{t-1} &=(S_{FiT}[1]S_i[1],\ldots,S_{FiT}[t-1]S_i[t-1]).
\end{align}
We set
\begin{align}
\label{Eq:SnSFn}
S[t] &\overset{\triangle}= \left( S_1[t], S_2[t] \right), \nonumber \\
S_{F1}[t] &\overset{\triangle}= \left( S_{F1T}[t], S_{F21}[t] \right), \nonumber \\
S_{F2}[t] &\overset{\triangle}= \left( S_{F2T}[t], S_{F12}[t] \right), \nonumber \\
S_F[t] &\overset{\triangle}= \left( S_{F1T}[t], S_{F12}[t], S_{F21}[t], S_{F2T}[t] \right).
\end{align}
We also define $S^t = \left( S[\ell] \right)_{\ell = 1}^t$, $S_{Fi}^t = \left( S_{Fi}[\ell] \right)_{\ell = 1}^t$, $i=1,2$, and $S_F^t = \left( S_F[\ell] \right)_{\ell = 1}^t$. Basically, at time instant $t$: $S[t]$ is the forward channel state information; $S_{Fi}[t]$ is the feedback state from receiver $i$ to the transmitter, and from receiver $\bar{i}$ to receiver $i$; and $S_F[t]$ is the entire feedback state information.


Each receiver $\msf{Rx}_i$, $i=1,2$, uses a decoding\footnote{In the initial results of this work~\cite{ITW-IFB} there is typo in the available information at each receiver. This typo does not affect the overall results and the edited version is available online at link provided with reference~\cite{ITW-IFB}.} function
\begin{align}
\label{Eq:DecodingFunction}
\varphi_{i,n}\left( Y_i^n, S_i^n, S_{Fi}^n, \lbp S_{F{\bar{i}}i}S_{\bar{i}} \rbp^{n} \right),
\end{align}
to get an estimate $\widehat{W}_i$ of $W_i$. An error occurs whenever $\widehat{W}_i \neq W_i$. The average probability of error is given by
\begin{align}
\lambda_{i,n} = \mathbb{E}[P(\widehat{W}_i \neq W_i)],
\end{align}
where the expectation is taken with respect to the random choice of the transmitted messages.

We say that a rate pair $(R_1,R_2)$ is achievable if there exists a block encoder at the transmitter, and a block decoder at each receiver, such that $\lambda_{i,n}$ goes to zero as the block length $n$ goes to infinity. The capacity region, $\mathcal{C}$, is the closure of the set of achievable rate pairs.

\begin{remark} [A note on available CSIR] \label{Remark:CSIR}
In this paper, we make certain assumptions on the available channel state information at the receiver(s) (CSIR). Consider ${\sf Rx}_1$, then, the CSIR as noted in \eqref{Eq:DecodingFunction} and depicted in Fig.~\ref{Fig:BC_IFB}(b), includes its corresponding channel state information $S_1[t]$; the feedback state from receiver $2$ to receiver $1$, \emph{i.e.} $S_{F\bar{i}i}[t]$ and $S_{F\bar{i}i}[t]S_{\bar{i}}[t]$; and the feedback state from receiver $1$ to the transmitter $S_{F1T}[t]$. The first two assumptions are justified as ${\sf Rx}_1$ is at the receiver end of those links. However, $S_{F1T}[t]$ is the link from receiver $1$ to the transmitter and the availability of this knowledge at ${\sf Rx}_1$ needs further justification. Erasure BCs typically capture packet networks where in the forward channel, packets containing thousands of bits are transmitted, yet feedback signals are individual bits. Extending our results to packet erasure BCs is straightforward, and thus, it is reasonable to assume the transmitter can provide $S_{F1T}[t]$ to ${\sf Rx}_1$ with negligible overhead. On the other hand, if we truly focus on a BC in which forward and feedback signals are all individual bits, then, this overhead is not negligible and the problem becomes more complicated.
\end{remark}

%% file: Main_IFB_symmetric.tex
The following theorem establishes an outer bound on the capacity region of the two-user erasure BC with intermittent feedback.

\begin{theorem}
\label{THM:Capacity_IFB}
The capacity region, $\mathcal{C}$, of the two-user erasure BC with intermittent feedback as described in Section~\ref{Section:Problem_IFB} is included in:
\begin{equation}
\label{Eq:Bounds_IFB}
\mathcal{C}_{\mathrm{out}} =
\left\{ \begin{array}{ll}
\hspace{-1.5mm} \left( R_1, R_2 \right) \left| \parbox[c][3em][c]{0.25\textwidth} {
$R_1 + \beta_2 R_2 \leq \beta_2 \left( 1 - \delta_2 \right)$\\
$\beta_1 R_1 + R_2 \leq \beta_1 \left( 1 - \delta_1 \right)$
} \right. \end{array} \right\}
\end{equation}
where for $i=1,2,$
\begin{align}
\label{Eq:Beta_IFB}
\beta_i = \frac{\delta_{FF} \left( 1 - \min_j \delta_j \right) + \left( 1- \delta_{FF} \right) \left( 1 - \delta_{12} \right)}{\left( 1 - \delta_i \right)},
\end{align}
and $\delta_{FF}$ is defined in \eqref{Eq:deltadefinitions}.
\end{theorem}


We derive the outer-bounds in two steps. First, we create a modified BC in which forward links are correlated across users when both feedback links to the transmitter are erased, \emph{i.e.}
\begin{align}
S_{F1T}[t] =0 \text{~and~} S_{F2T}[t] =0.
\end{align}
We show that the capacity region of this modified problem includes that of the original problem we are interested in. Next, we derive the outer-bounds for this modified problem which in turn serve as outer-bounds on $\mathcal{C}$. The details are provided in Section~\ref{Section:Converse_IFB}.

An interesting observation is that individual feedback erasure probabilities do not appear in these outer bounds, but rather these bounds are governed by the probability of missing both feedback links at the transmitter, that is, $\delta_{FF}$. In other words, as long as one receiver is successful in delivering its CSI to the transmitter, the outer-bound region in Theorem~\ref{THM:Capacity_IFB} matches the one with global delayed CSI. This observation is in agreement with our earlier result~\cite{ISIT19sclin} where  we showed that the capacity region of the two-user erasure broadcast channel with global delayed CSI can be achieved with single-user delayed CSI only.

{\begin{remark}[Comparison to the early work of Sato~\cite{sato1977two}]
In the information theory literature, it is known that the capacity of multi-terminal channels remains intact as long as the marginal distributions at the receivers remain the same~\cite{sato1977two}. To apply this result, one must first show the multi-letter marginal distributions remain intact under the induced correlation and the assumptions of the problem and then, find ``the worst-case'' correlation to obtain the tightest outer-bounds. Unfortunately, existing results in this direction do not immediately extend to the current work. The reason is the presence of feedback (unlike the no-feedback model of~\cite{sato1977two}) creates dependency on the CSI for the transmit signal across time, affecting the marginal distributions at the receivers. In other words, even satisfying the first step becomes a daunting task, not to mention finding the worst-case correlation. We were able to use the results of~\cite{sato1977two} in~\cite{ITW-IFB} where the intermittent feedback had a particular structure under which delayed CSI is either shared with both or neither of the other nodes. For the more general feedback structure assumed in this work, we present an alternative argument to demonstrate how and under what conditions, correlation among forward channel may be induced without shrinking the capacity region. 
\end{remark}}

Next, we show that under certain conditions, the outer-bound region $\mathcal{C}_{\mathrm{out}}$ can be achieved. These conditions were reported in~\cite{ITW-IFB} with shortened proofs, and in Section~\ref{Section:Achievability_IFB}, we present the detailed proof of the results. We then state a new theorem that provides an inner-bound for a new case compared to our earlier work.
\begin{theorem}
\label{THM:Ach_IFB}
The outer-bound region, $\mathcal{C}_{\mathrm{out}}$, on the capacity region of the two-user erasure BC with intermittent feedback as given in Theorem~\ref{THM:Capacity_IFB}, equals the capacity region, $\mathcal{C}$, when:
\begin{enumerate}
\item $\delta_{F1} \delta_{F2}= 0$, or
\item $S_{FiT}[t] = S_{Fi{\bar{i}}}[t]$, $i=1,2$,  $\delta_1 = \delta_2$, $\delta_{F1} = \delta_{F2}$, and $\Pr\left( S_{F1T}[t] \neq S_{F2T}[t] \right) = 0$.
\end{enumerate}
\end{theorem}


To prove the achievability, we propose a transmission strategy that has a recursive form. The first iteration has three phases which resemble the three phase communication of two-user BC with global feedback~\cite{GeorgiadisDetDelayedBC,Wang_12,GatzianasGeorgiadis_13}. After these three phases, we create recycled bits that we feed to the same transmission protocol as the recursive step. We show that the achievable rate matches the outer bounds of Theorem~\ref{THM:Capacity_IFB}. The details are provided in Section~\ref{Section:Achievability_IFB}. 

\begin{figure}[!ht]
\centering
\includegraphics[width = .4\columnwidth]{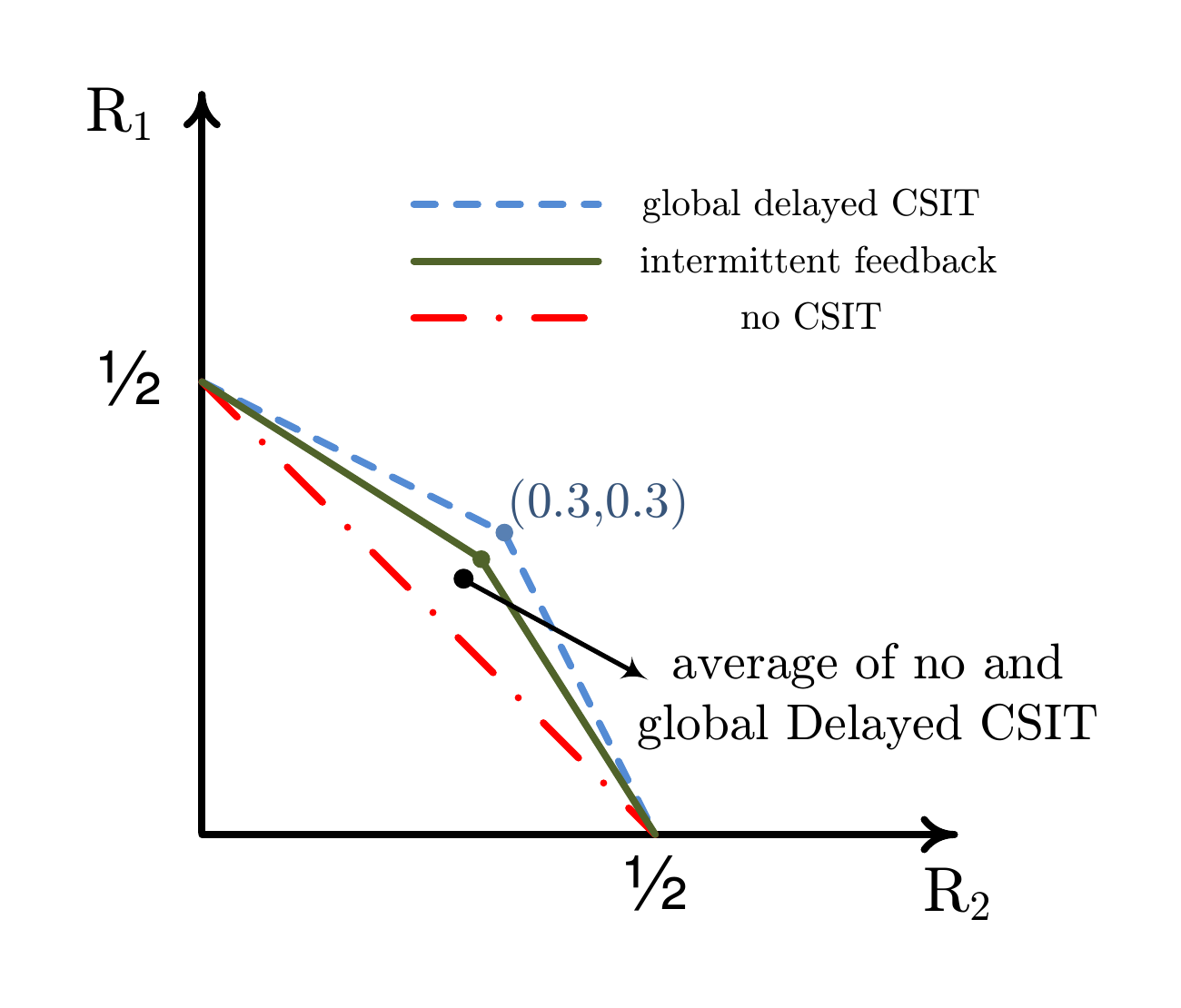}
\caption{The maximum achievable sum-rate of the erasure BC with intermittent feedback for parameters given in (\ref{Eq:ParametersExample}) is $5/9$ which is greater than the average of  the no feedback and the global feedback scenarios.\label{Fig:Average}}
\end{figure}

The above two theorems demonstrate how the capacity region degrades as the quality of feedback channel diminishes. However, an interesting observation is that the capacity region of the two-user erasure BC with intermittent feedback is larger than the average of the one with no feedback and the one with global feedback. To clarify this, consider an example in which feedback links from each receiver have the same erasure probability and are fully correlated (\emph{i.e.} are on and off together), and we have
\begin{align}
\label{Eq:ParametersExample}
& \delta_1 = \delta_2 = 0.5, \qquad \delta_{12} = 0.25, \nonumber \\
& \delta_{F1} = \delta_{F2} = \delta_{FF} = 0.5.
\end{align}
Note that this example falls under Case 2 of Theorem~\ref{THM:Ach_IFB}, and hence, the capacity region is characterized by the outer-bound region in Theorem~\ref{THM:Capacity_IFB}. For these parameters, the maximum achievable sum-rate without feedback is $0.5$, while that with global delayed CSIT is $0.6$ as depicted in Fig.~\ref{Fig:Average}. Moreover, half of the times at least one feedback link is active, that is, $\delta_{FF} = 0.5$. Averaging the maximum achievable sum-rates of the no feedback and the global feedback scenarios gives us $0.55$. Interestingly, the maximum achievable sum-rate of this problem with intermittent feedback is $5/9$ which is greater than $0.55$, the sum-rate achieved by time sharing (see Fig.~\ref{Fig:Average} for illustration). The intuition is as follows. For the no feedback case with $\delta_1 = \delta_2$, both receivers can decode both messages since the two receivers are stochastically equivalent. However, as we will show in Section~\ref{Section:Achievability_IFB}, our recursive transmission protocol efficiently exploits the available feedback, and prevents the suboptimal decoding of messages by both receivers in the naive time-sharing scheme.

Finally, we present an inner-bound when channel conditions are more relaxed compared to those in Theorem~\ref{THM:Ach_IFB} and we quantify the gap to the outer-bounds. The significance of this case is two-fold: (1) feedback links from the two receivers are not fully correlated and can act arbitrarily with respect to each other; (2) it provides an interesting connection to a well-known problem, namely Blind Index Coding (BIC)~\cite{kao2016blind} as we discuss below and further in Sections~\ref{Section:Achievability_InnerBound_IFB} and~\ref{Section:Discussion_IFB}.
\begin{theorem}
\label{THM:InnerBound_IFB}
For the two-user erasure BC with intermittent feedback, when
\begin{align}
\label{Eq:ConditionsInner_IFB}
S_{FiT}[t] = S_{Fi\bar{i}}[t], i=1,2; \delta_1 = \delta_2 = \delta_{F1} = \delta_{F2} = \delta; \text{~and~} \delta_{FF} = \delta_{12} = \delta^2,
\end{align}
we can achieve the following corner point:
\begin{align}
R_1 = R_2 = \frac{\left( 1 - \delta^2 \right)}{2+\delta+\delta^3}.
\end{align}
\end{theorem}

First, we discuss the gap between the inner-bound of Theorem~\ref{THM:InnerBound_IFB} and the outer-bound of Theorem~\ref{THM:Capacity_IFB}. Note that corner points, $\left( R_1, R_2 \right) = \left( 1-\delta, 0 \right)$ and $\left( R_1, R_2 \right) = \left( 0, 1-\delta \right)$ are trivially achievable. The maximum sum-rate, \emph{i.e.}
\begin{align}
\max_{R_1,R_2 \in \mathcal{C}_{\mathrm{out}}} R_1 + R_2,
\end{align}
from Theorem~\ref{THM:Capacity_IFB} under the conditions of Theorem~\ref{THM:InnerBound_IFB} in \eqref{Eq:ConditionsInner_IFB} is
\begin{align}
R_1 = R_2 = \frac{\left( 1 + \delta - \delta^3 \right) \left( 1 - \delta \right)}{2+\delta-\delta^3}.
\end{align}

\begin{figure}[!ht]
\centering
\includegraphics[width = .65\columnwidth]{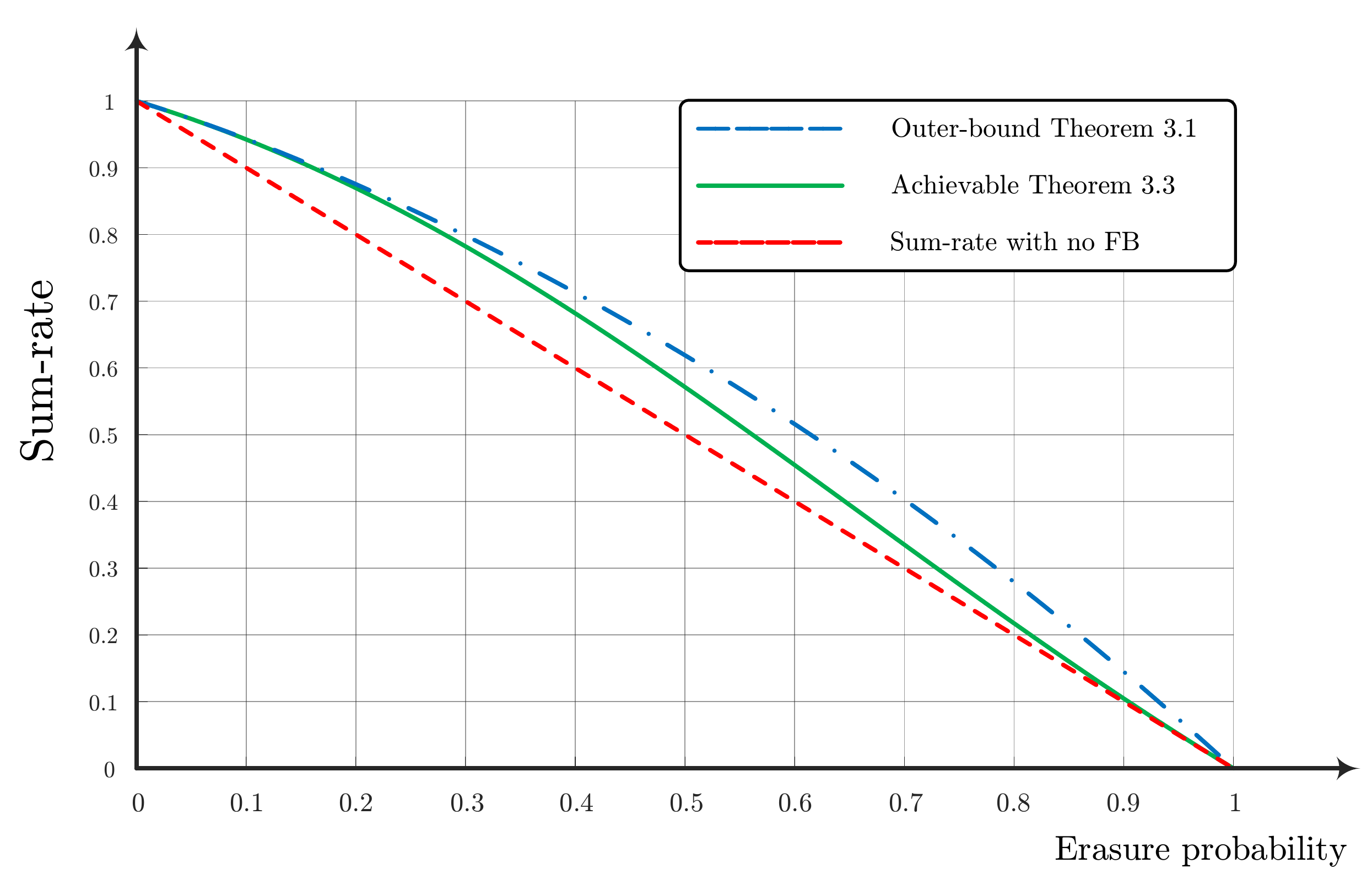}
\caption{Comparing the sum-rate inner-bound of Theorem~\ref{THM:InnerBound_IFB} to the outer-bounds of Theorem~\ref{THM:Capacity_IFB} when channel is described by \eqref{Eq:ConditionsInner_IFB}. The red dashed line is the sum-capacity of this problem with no channel feedback and is included as a reference.\label{Fig:InnerOuterIFB}}
\end{figure}

Figure~\ref{Fig:InnerOuterIFB} depicts these inner and outer bounds as well as the sum-capacity of this problem with no channel feedback as a reference. The proof of Theorem~\ref{THM:InnerBound_IFB} is based on an initial three-phase communication followed by a simple step where we sub-optimally ignore the feedback. This seemingly strange decision is rooted in the fact that, as we will discuss in further detail in Section~\ref{Section:Achievability_InnerBound_IFB}, finding the optimal solution requires solving the Blind Index Coding problem with intermittent feedback. However, Blind Index Coding in erasure BCs is a complicated problem even without including feedback~\cite{kao2016blind}, let alone intermittent feedback. We believe our recursive strategies introduced in this paper help shed light on BIC problems.

%% file: Converse_IFB.tex
We derive the outer bounds in two steps. First, we introduce a modified BC in which channels are distributed as in the original BC of Section~\ref{Section:Problem_IFB} except for when both feedback signals to the transmitter are erased, \emph{i.e.} 
\begin{align}
\label{Eq:CorrelationCondition}
S_{F1T}[t] = 0 \text{~and~} S_{F2T}[t] = 0.
\end{align} 
We show that any outer bound on the capacity region of this modified channel will serve as an outer bound on $\mathcal{C}$, the capacity region of the BC introduced in Section~\ref{Section:Problem_IFB}. Then, we obtain the outer bounds for this modified BC.

\noindent \underline{{\bf Step 1}}: Without loss of generality, we assume $\dl_1 \ge \dl_2$, \emph{i.e.} $\msf{Rx}_2$ has a stronger channel\footnote{Receiver 2 having a stronger channel does not imply that receiver 1 is degraded.}. Consider a two-user erasure BC as defined in Section~\ref{Section:Problem_IFB} with the following modification. When both feedback signals to the transmitter are erased, , \emph{i.e.} the condition in (\ref{Eq:CorrelationCondition}) is satisfied, we have
\begin{align}
\label{Eq:Changes}
& \tilde{S}_1[t] = G[t]S_2[t], \nonumber \\
& \tilde{S}_2[t] = S_2[t],
\end{align}
where we use $\tilde{\cdot}$ to distinguish the modified BC from the original one, $\lbp G[t]\rbp$ denotes a Bernoulli $(1-\delta_1)/(1-\delta_2)$ process that is distributed i.i.d. over time and is independent of all other channel parameters. All other channel parameters remain unchanged.

\begin{figure}[!ht]
\centering
\subfigure[]{\includegraphics[width = .35\columnwidth]{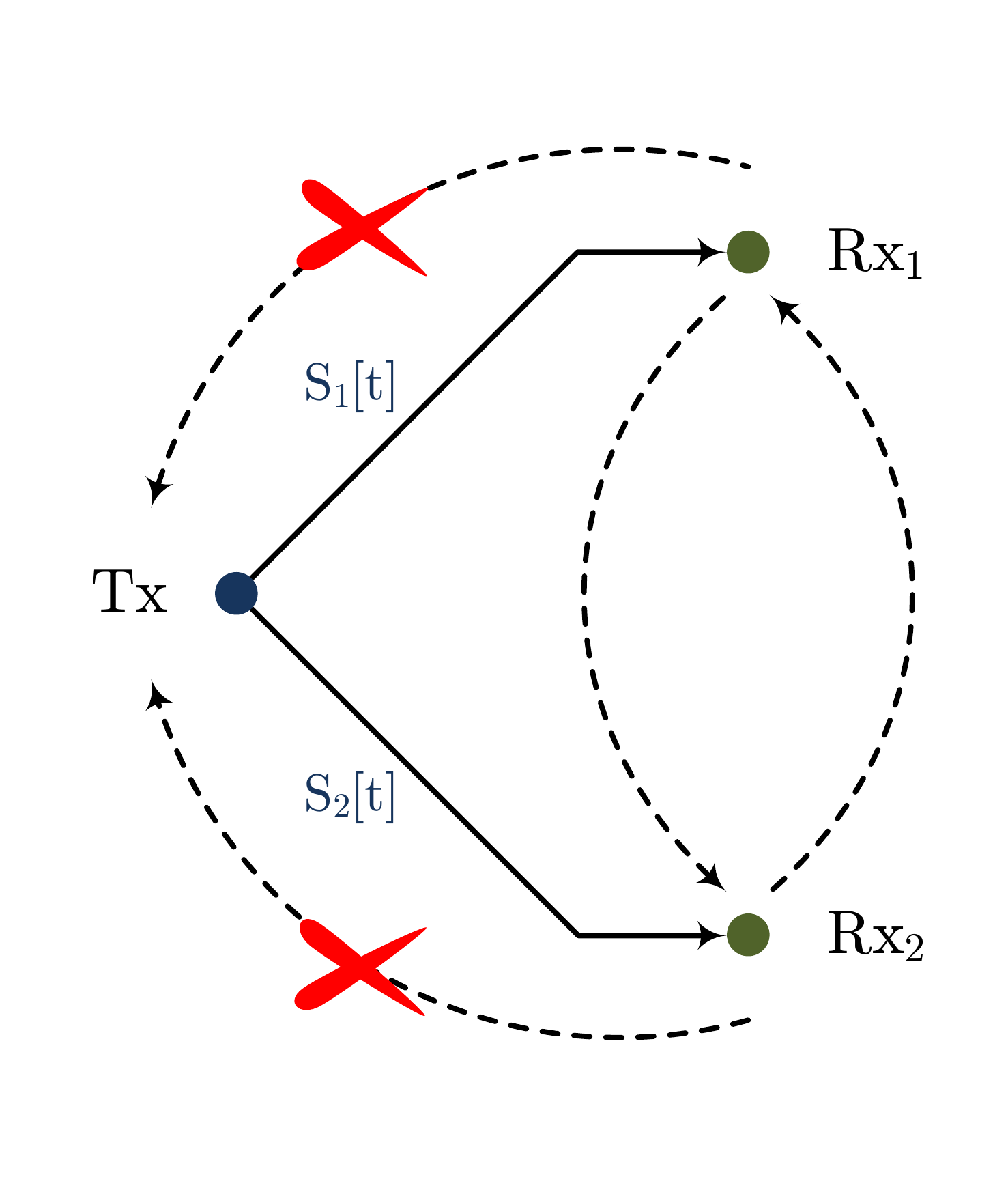}}
\hspace{.5in}
\subfigure[]{\includegraphics[width = .35\columnwidth]{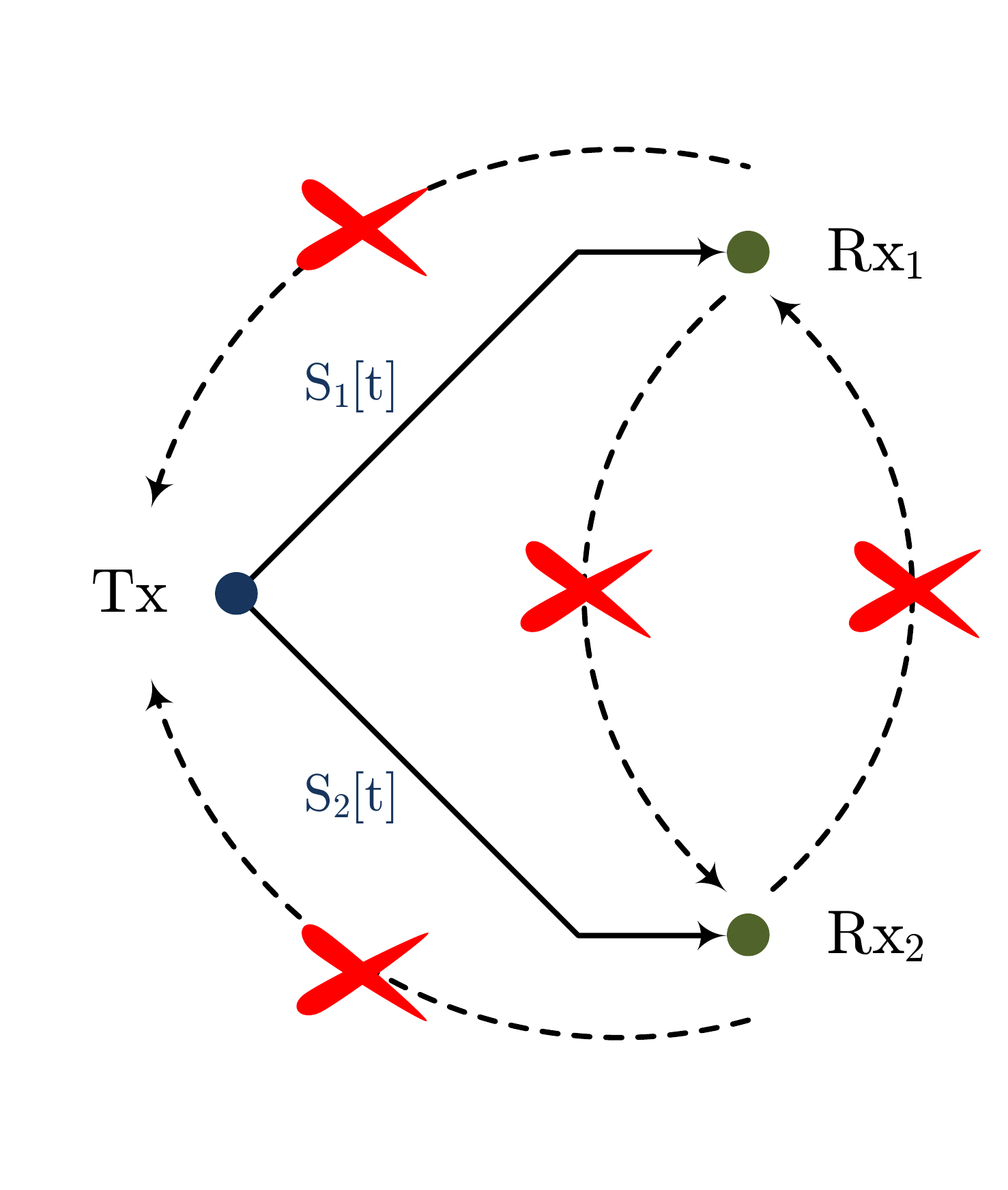}}
\caption{(a) When both feedback links to the transmitter are erased, we introduce correlation in the forward channel; (b) In our earlier work~\cite{ITW-IFB}, correlation was introduced when all feedback links were erased.}\label{Fig:FBCategories}
\end{figure}

Since here we focus on deriving outer-bounds, we further enhance the channels by ensuring $S_{F12}[t] = S_{F21}[t] = 1$ as well as $\tilde{S}_{F12}[t] = \tilde{S}_{F21}[t] = 1$. We note that in our earlier work~\cite{ITW-IFB}, we required $S_{12}[t] = S_{21}[t] = 0$ on top of $S_{F1T}[t] = S_{F2T}[t] = 0$ as depicted in Fig.~\ref{Fig:FBCategories}(b).

\begin{claim}
\label{Claim_modified}
The capacity region of the modified two-user erasure BC with intermittent feedback as described above includes that of the two-user erasure BC with intermittent feedback described in Section~\ref{Section:Problem_IFB}.
\end{claim}

{
\begin{remark}
The converse proof in~\cite{ITW-IFB} for the feedback structure  of Fig.~\ref{Fig:FBCategories}(b) is based on proving the equivalence of marginal distributions at the receivers in the original and the modified BCs. In other words, for the specific intermittent feedback model of~\cite{ITW-IFB} where delayed CSI is either shared with both or neither of the other nodes, we could use the results of~\cite{sato1977two}. However, in our more general setting, we take a direct approach by showing the mutual information outer-bounds on individual rates remain unchanged under the modification of \eqref{Eq:Changes}. 
\end{remark}}

\noindent {\it Proof of Claim~\ref{Claim_modified}:}
If at time $t$, the condition in (\ref{Eq:CorrelationCondition}) is satisfied, \emph{i.e.} $S_{F1T}[t] = 0$ and $S_{F2T}[t] = 0$ as in Fig.~\ref{Fig:FBCategories}(a), then the transmitter will never learn the values of $S_1[t]$ and $S_2[t]$.  

Now, suppose in the original problem, rate-tuple $(R_1, R_2)$ is achievable, and $W_1$ and $W_2$ are encoded as $X^n$. Moreover, suppose the condition in (\ref{Eq:CorrelationCondition}) is satisfied for the first time at time instant $t > 0$ (the condition is irrelevant for $t=0$). In other words, up until time instant $t-1$, the modified and the original channels are identical. To simplify the notation, we define
\begin{align}
A_1[t] &\overset{\triangle}= \left( Y_1[t], S_1[t], S_{F1}[t], S_{F21}[t] S_2[t] \right) \nonumber \\
A_2[t] &\overset{\triangle}= \left( Y_2[t], S_2[t], S_{F2}[t], S_{F12}[t] S_1[t] \right).
\end{align}
Also, for consistency, we set $A_i[0]$ deterministically to zero values, $i = 1,2$. Then, we have
\begin{align}
\label{Eq:ExpandingMIRx1}
n (R_1-\epsilon_n) &\overset{\mathrm{Fano}}\leq I\left( W_1; A_1^n \right) \nonumber \\
&= \sum_{\ell=1}^{n}{I\left( W_1; A_1[\ell] | A_1^{\ell-1} \right)} \nonumber \\
&= \sum_{\ell=1}^{t-1}{I\left( W_1; A_1[\ell] | A_1^{\ell-1} \right)} \nonumber \\
&~+I\left( W_1; \underbrace{Y_1[t], S_1[t], S_{F1}[t], S_{F21}[t] S_2[t]}_{=~A_1[t]} | A_1^{t-1} \right) \nonumber \\
&~+\sum_{\ell=t+1}^{n}{I\left( W_1; A_1[\ell] | A_1^{\ell-1} \right)}.
\end{align} 

As mentioned above, suppose the condition in (\ref{Eq:CorrelationCondition}) is satisfied for the first time at time instant $t > 0$, meaning that $S_{F1T}[t] = S_{F2T}[t] = 0$ and no feedback is provided to the transmitter. Then, since we enhanced the channel by setting $S_{F12}[t] = S_{F21}[t] = 1$, for receiver ${\sf Rx}_1$, we have
\begin{align}
\label{Eq:TimetCategory1Rx1}
&I\left( W_1; A_1[t] | A_1^{t-1} \right) \nonumber \\
&= I\left( W_1; Y_1[t], S_1[t], S_{F1}[t], S_{F21}[t] S_2[t] | A_1^{t-1} \right) \nonumber \\
&\overset{(a)}= I\left( W_1; Y_1[t], S_1[t], S_2[t] | A_1^{t-1} \right) \nonumber \\
&= \underbrace{I\left( W_1; S_1[t], S_2[t] | A_1^{t-1} \right)}_{=~0} + I\left( W_1; Y_1[t] | S_1[t], S_2[t], A_1^{t-1} \right) \nonumber \\
&\overset{(b)}= I\left( W_1; Y_1[t] | S_1[t], S_2[t], A_1^{t-1} \right) \nonumber \\
&\overset{(c)}= I\left( W_1; Y_1[t] | S_1[t], A_1^{t-1} \right) \nonumber \\
&= I\left( W_1; Y_1[t] | S_1[t], A_1^{t-1} \right) + \underbrace{I\left( W_1; S_1[t] | A_1^{t-1} \right)}_{=~0} \nonumber \\
&= I\left( W_1; Y_1[t], S_1[t] | A_1^{t-1} \right),
\end{align}
where $(a)$ holds since at time $t$ as discussed above, $S_{F1T}[t] = S_{F2T}[t] = 0$ and $S_{F12}[t] = S_{F21}[t] = 1$; $(b)$ holds since channels at time $t$ are independent of the messages and previous channel realizations;
$(c)$ follows from the fact the $X[t]$ is independent of $S_2[t]$, and thus, so is $Y_1[t]$. Moreover, 
\begin{align}
\label{Eq:Timetplus1Category1Rx1}
& I\left( W_1; A_1[t+1] | A_1^{t} \right) \nonumber \\
&= I\left( W_1; A_1[t+1]| Y_1[t], S_1[t], S_{F1}[t], S_{F21}[t] S_2[t], A_1^{t-1} \right) \nonumber \\
&\overset{(a)}= I\left( W_1; A_1[t+1]| Y_1[t], S_1[t], S_2[t], A_1^{t-1} \right) \nonumber \\
&\overset{(b)}= I\left( W_1; A_1[t+1]| Y_1[t], S_1[t], A_1^{t-1} \right),
\end{align}
where $(a)$ holds since at time $t$ as discussed above, $S_{F1T}[t] = S_{F2T}[t] = 0$ and $S_{F12}[t] = S_{F21}[t] = 1$; $(b)$ holds since $X[t]$ and $Y_1[t]$ are independent of $S_2[t]$,  and since $S_{F1T}[t] = S_{F2T}[t] = 0$, $X[t+1]$ is also independent of $S_2[t]$. 

Comparing \eqref{Eq:TimetCategory1Rx1} and \eqref{Eq:Timetplus1Category1Rx1}, we conclude that when condition \eqref{Eq:CorrelationCondition} is met, we can remove $S_2[t]$ from $A_1[t]$ but still keep \eqref{Eq:ExpandingMIRx1} the same.

Similar arguments hold for receiver ${\sf Rx}_2$:
\begin{align}
\label{Eq:TimetCategory1Rx2}
&I\left( W_2; A_2[t] | A_2^{t-1} \right) \nonumber \\
&= I\left( W_2; Y_2[t], S_2[t], S_{F2}[t], S_{F12}[t] S_1[t] | A_2^{t-1} \right) \nonumber \\
&\overset{(a)}= I\left( W_2; Y_2[t], S_1[t], S_1[t] | A_2^{t-1} \right) \nonumber \\
&= \underbrace{I\left( W_2; S_1[t], S_2[t] | A_2^{t-1} \right)}_{=~0} + I\left( W_2; Y_2[t] | S_1[t], S_2[t], A_2^{t-1} \right) \nonumber \\
&\overset{(b)}= I\left( W_2; Y_2[t] | S_1[t], S_2[t], A_2^{t-1} \right) \nonumber \\
&\overset{(c)}= I\left( W_2; Y_2[t] | S_2[t], A_2^{t-1} \right) \nonumber \\
&= I\left( W_2; Y_2[t] | S_2[t], A_2^{t-1} \right) + \underbrace{I\left( W_2; S_2[t] | A_2^{t-1} \right)}_{=~0} \nonumber \\
&= I\left( W_2; Y_2[t], S_2[t] | A_2^{t-1} \right),
\end{align}
where $(a)$ holds since $S_{F1T}[t] = S_{F2T}[t] = 0$ and we enhanced the feedback links by setting $S_{F12}[t] = S_{F21}[t] = 1$; $(b)$ holds due to the independence of the messages and previous signals from the channel gains at time $t$; $(c)$ follows from the fact the $X[t]$ is independent of $S_1[t]$, and thus, so is $Y_2[t]$. We also have
\begin{align}
\label{Eq:Timetplus1Category1Rx2}
& I\left( W_2; A_2[t+1] | A_2^{t} \right) \nonumber \\
&= I\left( W_2; A_2[t+1]| Y_2[t], S_2[t], S_{F2}[t], S_{F12}[t] S_1[t], A_2^{t-1} \right) \nonumber \\
&\overset{(a)}= I\left( W_2; A_2[t+1]| Y_2[t], S_1[t], S_2[t], A_2^{t-1} \right) \nonumber \\
&\overset{(b)}= I\left( W_2; A_2[t+1]| Y_2[t], S_2[t], A_2^{t-1} \right),
\end{align}
where $(a)$ holds since at time $t$, we have $S_{F1T}[t] = S_{F2T}[t] = 0$ and $S_{F12}[t] = S_{F21}[t] = 1$; $(b)$ holds since $X[t]$ and $Y_2[t]$ are independent of $S_1[t]$,  and since we have $S_{F1T}[t] = S_{F2T}[t] = 0$, then $X[t+1]$ is also independent of $S_1[t]$.

From the discussion above, we conclude that 
\begin{align}
\label{Eq:EuivalentMIRx1}
n (R_1-\epsilon_n) &\overset{\mathrm{Fano}}\leq I\left( W_1; A_1^n \right) \nonumber \\
&= I\left( W_1; Y_1^n, S_1^n, S_{F1}^n, \{ S_{F21} S_2 \}^n \right) \nonumber \\
&= I\left( W_1; Y_1^n, S_1^n, S_{F1}^n, \{ S_{F21} S_2 \}^{n \setminus \mathrm{C}} \right),
\end{align} 
where $\cdot^{n \setminus \mathrm{C}}$ represents the length-$n$ vector with elements removed when $S_{F1T}[t] = 0$ and $S_{F2T}[t] = 0$. In other words, $ \{ S_{F21} S_2 \}^{n \setminus \mathrm{C}}=\{ (S_{F1T} \vee S_{F2T})S_{F21} S_2 \}^{n}$. We further conclude that
\begin{align}
\label{Eq:EuivalentMIRx2}
n (R_2-\epsilon_n) &\overset{\mathrm{Fano}}\leq I\left( W_2; A_2^n \right) \nonumber \\
&= I\left( W_2; Y_2^n, S_2^n, S_{F2}^n, \{ S_{F12} S_1 \}^n \right) \nonumber \\
&= I\left( W_2; Y_2^n, S_2^n, S_{F2}^n, \{ S_{F12} S_1 \}^{n \setminus \mathrm{C}} \right).
\end{align}

Going back to the outer-bound on individual rates in \eqref{Eq:ExpandingMIRx1}, since we assumed up until time instant $t-1$, the modified and the original channels are identical, we have
\begin{align}
\label{Eq:PriorTot}
\sum_{\ell=1}^{t-1}{I\left( W_1; A_1[\ell] | A_1^{\ell-1} \right)} = \sum_{\ell=1}^{t-1}{I\left( W_1; \tilde{A}_1[\ell] | \tilde{A}_1^{\ell-1} \right)},
\end{align}
where $\tilde{\cdot}$ denotes the signals in the modified channel, and note that here we used
\begin{align}
\label{Eq:sameX}
\tilde{Y}_1[t] = \tilde{S}_1[t]X[t].
\end{align}
Moreover, at time $t$, when (\ref{Eq:CorrelationCondition}) is satisfied for the first time, we have
\begin{align}
 \label{Eq:TimetCategory1Rx1Mod}
&I\left( W_1; A_1[t] | A_1^{t-1} \right) \nonumber \\
&\overset{\eqref{Eq:TimetCategory1Rx1}}= I\left( W_1; Y_1[t] | S_1[t], A_1^{t-1} \right) \nonumber \\
&\overset{(a)}= I\left( W_1; \tilde{Y}_1[t] | \tilde{S}_1[t], \tilde{A}_1^{t-1} \right) \nonumber \\
&\overset{(b)}= I\left( W_1; \tilde{Y}_1[t], \tilde{S}_1[t] | \tilde{A}_1^{t-1} \right),
\end{align}
where $(a)$ holds since at time $t$, $X[t]$ is independent of $S_1[t]$, while $\tilde{Y}_1[t]$, $\tilde{S}_1[t]$ and $\tilde{A}_1^{t-1}$ are statistically the same as $Y_1[t]$, $S_1[t]$, and $A_1^{t-1}$, respectively; $(b)$ holds since channels at time $t$ are independent of the messages and previous channel realizations. 

Finally, since the transmitter never learns the values of $S_1[t]$ and $S_2[t]$ in the original channel, we have
\begin{align}
 \label{Eq:AfterTimetCategory1Rx1Mod}
&I\left( W_1; A_1[t+1] | A_1^{t} \right) \nonumber \\
&\overset{\eqref{Eq:Timetplus1Category1Rx1}}= I\left( W_1; A_1[t+1]| Y_1[t], S_1[t], A_1^{t-1} \right) \nonumber \\
&\overset{(a)}= I\left( W_1; \tilde{A}_1[t+1]| \tilde{Y}_1[t], \tilde{S}_1[t], \tilde{A}_1^{t-1} \right)
\end{align}
where $(a)$ holds since $X[t+1]$ is independent of channel realizations at time $S_{1}[t]$ and $S_2[t]$, and further, $\tilde{Y}_1[t]$, $\tilde{S}_1[t]$ and $\tilde{A}_1^{t-1}$ are statistically the same as $Y_1[t]$, $S_1[t]$, and $A_1^{t-1}$, respectively.


We note that the only difference between the modified channel and the original channel is the correlation between the forward links in the modified channel, described in \eqref{Eq:Changes}, and all other parameters are the same for both channels. Now, from \eqref{Eq:PriorTot}, \eqref{Eq:TimetCategory1Rx1Mod}, and \eqref{Eq:AfterTimetCategory1Rx1Mod}, we  obtain
\begin{align}
\label{Eq:ExpandingMIRx1Modified}
n (R_1-\epsilon_n) &\overset{\mathrm{Fano}}\leq I\left( W_1; A_1^n \right) \nonumber \\
&= I\left( W_1; Y_1^n, S_1^n, S_{F1}^n, \{ S_{F21} S_2 \}^{n \setminus \mathrm{C}} \right) \nonumber \\
&= I\left( W_1; \tilde{Y}_1^n, \tilde{S}_1^n, \tilde{S}_{F1}^n, \{ \tilde{S}_{F21} \tilde{S}_2 \}^{n \setminus \mathrm{C}} \right),
\end{align}
where as mentioned earlier, we used
\begin{align}
\tilde{Y}_1[t] = \tilde{S}_1[t]X[t].
\end{align}
We have a similar result for ${\sf Rx}_2$:
\begin{align}
\label{Eq:ExpandingMIRx2Modified}
n (R_2-\epsilon_n) &\overset{\mathrm{Fano}}\leq I\left( W_2; A_2^n \right) \nonumber \\
&= I\left( W_2; Y_2^n, S_2^n, S_{F2}^n, \{ S_{F12} S_1 \}^{n \setminus C} \right) \nonumber \\
&= I\left( W_2; \tilde{Y}_2^n, \tilde{S}_2^n, \tilde{S}_{F2}^n, \{ \tilde{S}_{F12} \tilde{S}_1 \}^{n \setminus C} \right).
\end{align}

From \eqref{Eq:ExpandingMIRx1Modified} and \eqref{Eq:ExpandingMIRx2Modified}, we conclude that if rate-tuple $(R_1, R_2)$ is achievable in the original network, then, it is also achievable in the modified channel.  This in turn implies that the capacity region of the modified two-user erasure BC with intermittent feedback as described above includes that of the two-user erasure BC with intermittent feedback described in Section~\ref{Section:Problem_IFB}. \hfill{\hbox{$\blacksquare$}}

Claim~\ref{Claim_modified} implies that any outer bound on the capacity region of the modified BC serves as an outer bound on that of the problem described in Section~\ref{Section:Problem_IFB}. In Step~2, we derive the outer bounds on the capacity region of the modified BC.  

\noindent \underline{{\bf Step 2}}: The arguments below are presented for the modified BC, but with a slight abuse of notation, we drop the $\tilde{\cdot}$ for deriving the outer-bounds. Suppose rate-tuple $(R_1, R_2)$ is achievable in the modified BC. To obtain the first outer-bound in \eqref{Eq:Bounds_IFB}, let
\begin{align}
\label{Eq:Beta2_IFB}
\beta_2 = \frac{\delta_{FF} \left( 1 - \min_j \delta_j \right) + \left( 1- \delta_{FF} \right) \left( 1 - \delta_{12} \right)}{\left( 1 - \delta_2 \right)},
\end{align}
where as given in \eqref{Eq:deltadefinitions},
\begin{align}
\delta_{FF} \overset{\triangle}= P\{ S_{F1T}[t] = S_{F2T}[t] =0 \}.
\end{align}
By further giving $S^n_{F\bar{i}T}$ to $\msf{Rx}_i$ and additional $W_2$ to $\msf{Rx}_1$ in the modified BC, we have
\begin{align}
n & \left( R_1 + \beta_2 R_2 \right) = H\left(W_1\right) + \beta_2 H\left(W_2\right) \nonumber \\
& = H\left( W_1|W_2,S^n,S_F^n \right) + \beta_2 H\left( W_2|S^n,S_F^n \right) \nonumber \\
& \overset{\mathrm{Fano}}\leq I\left( W_1; Y_1^n|W_2,S^n,S_F^n \right) + \beta_2 I\left( W_2; Y_2^n|S^n,S_F^n \right) + n \epsilon_n \nonumber \\
& = H\left( Y_1^n|W_2,S^n,S_F^n \right) - \underbrace{H\left( Y_1^n|W_1,W_2,S^n,S_F^n \right)}_{=~0} \nonumber \\
&~+ \beta_2 H\left( Y_2^n|S^n,S_F^n \right) - \beta_2 H\left( Y_2^n|W_2,S^n,S_F^n \right) + n \epsilon_n \nonumber \\
& \overset{(a)}\leq \beta_2 H\left( Y_2^n|S^n,S_F^n \right) + n \epsilon_n \overset{(b)}\leq n \beta_2 (1-\delta_2) + n \epsilon_n,
\end{align}
where $\epsilon_n \rightarrow 0$ when $n \rightarrow \infty$; $S^n$ and $S_F^n$, as defined in \eqref{Eq:SnSFn}, encompass the forward and the feedback channels; $(a)$ follows from Claim~\ref{Claim_beta} below, and $(b)$ holds since $S_2[t]$ is Bernoulli $(1-\delta_2)$. The other outer-bound can be obtained similarly.

\begin{claim}
\label{Claim_beta}
For the two-user erasure BC with intermittent feedback as  described in Section~\ref{Section:Problem_IFB}, and for any input distribution, we have
\begin{align}
H\left( Y_1^n|W_2,S^n,S_F^n \right) - \beta_2 H\left( Y_2^n|W_2,S^n,S_F^n \right) \leq 0.
\end{align}
\end{claim}

\noindent {\it Proof of Claim~\ref{Claim_beta}:}
\begin{align}
&H\left( Y_2^n|W_2,S^n,S_F^n \right) \nonumber \\
& \overset{(a)}= \sum_{t=1}^{n}{H\left( Y_2[t]|Y_2^{t-1},W_2,S^t,S_F^t \right)} \nonumber \\
& \overset{(b)}= \sum_{t=1}^{n}{(1-\delta_2)H\left( X[t]|Y_2^{t-1},W_2,S_1[t], S_2[t]=1,S^{t-1},S_F^t \right)} \nonumber \\
& \overset{(c)}= \sum_{t=1}^{n}{(1-\delta_2)H\left( X[t]|Y_2^{t-1},W_2,S^{t},S_F^t \right)} \nonumber \\
& \overset{(d)}\ge \sum_{t=1}^{n}{(1-\delta_2)H\left( X[t]|Y_1^{t-1},Y_2^{t-1},W_2,S^{t},S_F^t \right)} \nonumber \\
& \overset{(e)}= \sum_{t=1}^{n}{\frac{(1-\delta_2)H\left( Y_1[t], Y_2[t] |Y_1^{t-1},Y_2^{t-1},W_2,S^{t},S_F^t \right)}{\delta_{FF} \left( 1 - \min_j \delta_j \right) + \left( 1- \delta_{FF} \right) \left( 1 - \delta_{12} \right)}} \nonumber \\
& \overset{(\ref{Eq:Beta2_IFB})}= \sum_{t=1}^{n}{\frac{1}{\beta_2}H\left( Y_1[t], Y_2[t] |Y_1^{t-1},Y_2^{t-1},W_2,S^{t},S_F^t \right)} \nonumber \\
& \overset{(f)}= \sum_{t=1}^{n}{\frac{1}{\beta_2}H\left( Y_1[t], Y_2[t] |Y_1^{t-1},Y_2^{t-1},W_2,S^{n},S_F^n \right)} \nonumber \\
& = {\frac{1}{\beta_2}H\left( Y_1^n, Y_2^n |W_2,S^{n},S_F^n \right)} \nonumber \\
&  \overset{(g)}\ge {\frac{1}{\beta_2}H\left( Y_1^n|W_2,S^{n},S_F^n \right)},
\end{align}
where $(a)$ follows since $X[t]$ is independent of future channel realizations and the channel gains are distributed as i.i.d. random variables over time, $(b)$ holds since $\Pr\left( S_2[t] = 1 \right) = (1-\delta_2)$,  $(c)$ follows from the same logic as step $(a)$, $(d)$ holds since conditioning reduces entropy, $(e)$ follows from the fact that $X[t]$ is independent from $S[t]$ and $S_F[t]$, and applying the total probability law as below
\begin{align}
& \Pr\left( \{ S_1[t] =  S_2[t] = 0 \}^{c} \right) = 1 - \Pr\left( S_1[t] = S_2[t] = 0 \right) \nonumber \\
& = 1 - \delta_{FF}\Pr\left( S_1[t] = S_2[t] = 0 | \{ S_{F1T}[t] = S_{F2T}[t] = 0 \} \right) \nonumber \\
&~- (1-\delta_{FF})\Pr\left( S_1[t] = S_2[t] = 0 | \{ S_{F1T}[t] = S_{F2T}[t] = 0 \}^{c} \right) \nonumber \\
& = 1 - \delta_{FF}  \min_j \delta_j - (1-\delta_{FF}) \delta_{12},
\end{align}
where \eqref{Eq:Changes} is used in the last equality; $(f)$ follows from the same logic as step $(a)$, and $(g)$ follows from the non-negativity of differential entropy. \hfill{\hbox{$\blacksquare$}}

%% file: Achievability_IFB_March2020.tex
We start from the achievability of Case~1, which is a review of our previous work \cite{ISIT19sclin}, and then present the new achievability for the other case.

\noindent {\bf Case~1}: If $\delta_{F1} = 0$ and $\delta_{F2} = 0$, then the capacity region of the two-user BC with intermittent FB is equivalent to that of global delayed CSI assumption (or ``DD'' assumption)~\cite{GeorgiadisDetDelayedBC}. Note that based on (\ref{Eq:deltadefinitions}), $\delta_{Fi} = 0$, $i=1,2$, means feedback links originating at ${\sf Rx}_i$ are always active and no erasure happens to the feedback signal of that receiver.

On the other hand, if one of the $\delta_{Fi}$'s is non-zero, then the capacity includes that of a two-user erasure BC with one-sided feedback (``DN'' assumption) for which we recently proved the capacity region coincides with that of DD assumption~\cite{ISIT19sclin}.

\begin{figure}[!ht]
\centering
\includegraphics[width = \columnwidth]{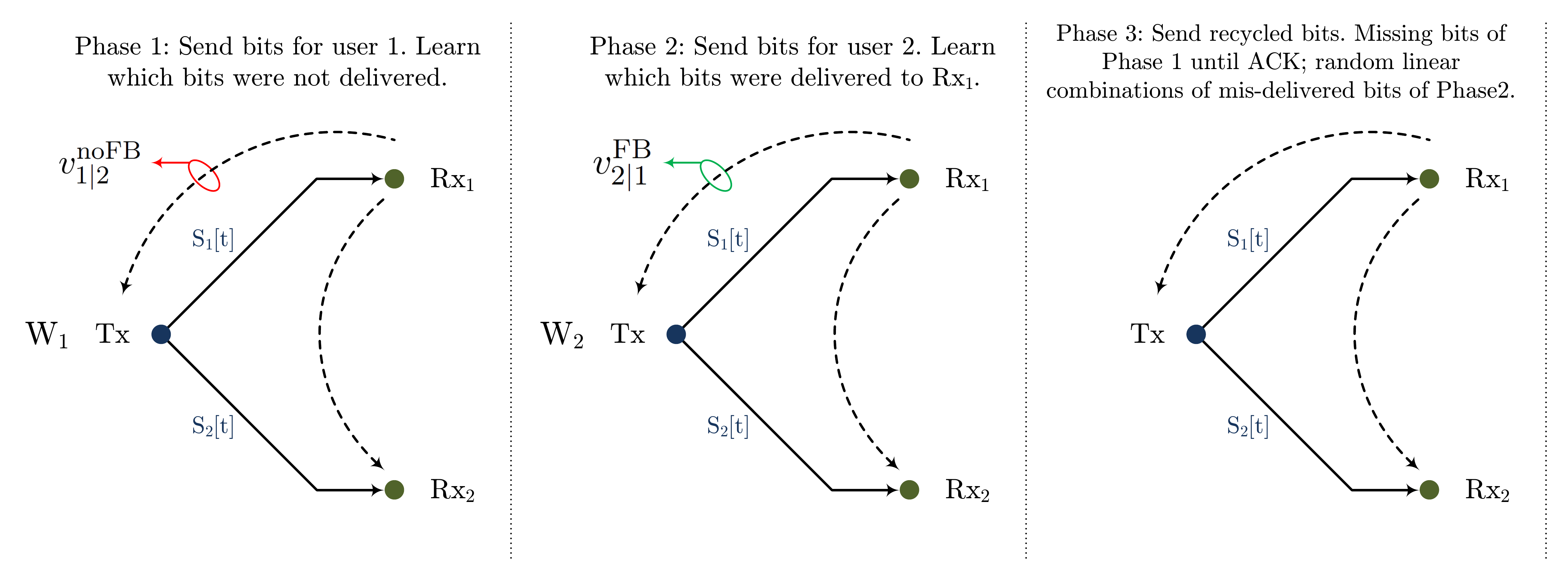}
\caption{A summary of the transmission protocol of~\cite{ISIT19sclin} with one-sided delayed channel state information (``DN'' assumption) for the achievability of Case 1 in Theorem~\ref{THM:Ach_IFB}.}
\label{Fig:DN}
\end{figure}

Here, we include a very brief overview of the achievability strategy under the DN assumption, both for better understanding the upcoming new achievability and for our discussion later in Section~\ref{Section:Discussion_IFB}. In the DN scenario, suppose only receiver 1 provides its feedback and receiver 2 is always silent. Furthermore, there is no erasure possibility and the CSI of receiver 1 is available to the other nodes with unit delay. The transmission strategy includes three phases as illustrated in Fig.~\ref{Fig:DN}. During the first phase, the transmitter communicates the bits intended for ${\sf Rx}_1$ and through the one-sided feedback learns $v_{1|2}^{\mathrm{noFB}}$, bits missing at ${\sf Rx}_1$. A fraction $(1-\delta_2)$ of these bits are available at ${\sf Rx}_2$. During the second phase, the transmitter communicates the bits intended for ${\sf Rx}_2$ and through the one-sided feedback learns $v_{2|1}^{\mathrm{FB}}$, coded bits delivered to the unintended receiver, \emph{i.e.} ${\sf Rx}_1$.  Note that $v_{1|2}^{\mathrm{noFB}}$  is of interest to ${\sf Rx}_1$ and \underline{partially} known at ${\sf Rx}_2$, where the ``$\mathrm{noFB}$'' superscript indicates lack of feedback from ${\sf Rx}_2$; also $v_{2|1}^{\mathrm{FB}}$, which is of interest to ${\sf Rx}_2$ and \underline{fully} known at ${\sf Rx}_1$. During the third phase, the transmitter sends out the XOR of a bit in $v_{1|2}^{\mathrm{noFB}}$ and a random linear combination of $v_{2|1}^{\mathrm{FB}}$. The bit in $v_{1|2}^{\mathrm{noFB}}$ is repeated as ARQ. That is, based on feedback, a new bit joins the XOR process only when the previous bit in $v_{1|2}^{\mathrm{noFB}}$ is delivered to the intended receiver ${\sf Rx}_1$. Remind that $v_{2|1}^{\mathrm{FB}}$ is fully known at ${\sf Rx}_1$, the interference from it can be removed from the delivered XOR. Note that unlike prior results, feedback is utilizes during all three phases of communications in order to achieve the DD capacity. Details of the rate analysis are presented in~\cite{ISIT19sclin}.


\noindent {\bf Case~2}: $\delta_{F1} = \delta_{F2} \overset{\triangle}= \delta_F$ and $\Pr\left( S_{F1T}[t] \neq S_{F2T}[t] \right) = 0$. This also implies that $\delta_{FF} = \delta_F$ since we also have $S_{F_iT}[t]=S_{F_{i\bar{i}}}[t], i=1,2$. Moreover, we have $\delta_1 = \delta_2  \overset{\triangle}= \delta$.

In this case, we can simplify some expressions as follows.
\begin{align}
\label{Def:A}
\beta_i = \frac{(1-\delta_{12})-\delta_F(\delta-\delta_{12})}{1-\delta} \overset{\triangle}= \frac{A}{1-\delta}.
\end{align}
Then, from Theorem~\ref{THM:Capacity_IFB}, we obtain the maximum sum-rate corner point:
\begin{align}
\label{Eq:TargetRates}
R_1 = R_2 = \frac{A}{1+\frac{A}{(1-\delta)}}.
\end{align}

\noindent {\bf Transmission Protocol}: The transmission strategy has a recursive format. We start with $m$ bits for each receiver. We start by sending the bits for each receiver and based on the available feedback, we create two sets of recycled equations. For the first set, feedback bits were available whereas for the second set, no feedback was available. We then send the XOR of the bits in first set. Whatever is left will be fed as the input to the transmission protocol again.

\noindent {\bf Phase~1}: The transmitter creates
\begin{align}
\frac{m}{1-\delta_{12}}
\end{align}
linearly independent equations of the $m$ bits for receiver $1$ and sends them out. For $(1-\delta_F)$ fraction of the time, feedback is available and
\begin{align}
\label{Eq:Rx1FB}
(1-\delta_F) \frac{\delta-\delta_{12}}{1-\delta_{12}} m
\end{align}
bits are available at ${\sf Rx}_2$ and needed at ${\sf Rx}_1$. Denote these bits by $v_{1|2}$. Moreover, for $\delta_F$ fraction of the time, no feedback is available. But due to the statistics of the channel, we know
\begin{align}
\label{Eq:Rx1noFB}
\delta_F \frac{\delta-\delta_{12}}{1-\delta_{12}} m = \frac{\delta_F}{1-\delta_F} \left| v_{1|2} \right|
\end{align}
bits are available at ${\sf Rx}_2$ and needed at ${\sf Rx}_1$. The transmitter creates the same number of linearly independent equations as (\ref{Eq:Rx1noFB}) from the transmitted bits $X[t]$ when feedback links were not available. Denote these equations by $v^{\mathrm{noFB}}_{1|2}$.

\begin{remark}
\label{remark:Bernstein}
To keep the description of the protocol simple, we use the expected value of the number of bits in different states, \emph{e.g.}, (\ref{Eq:Rx1FB}) and (\ref{Eq:Rx1noFB}). A more precise statement would use a concentration theorem result such as the Bernstein inequality to show the omitted terms do not affect the overall result and the achievable rates. A detailed example of such calculations can be found in Section V.B of~\cite{vahid2016two}. Moreover, when talking about the number of bits or equations, we are limited to integer numbers. If a ratio is not an integer number, we can use $\lceil \cdot \rceil$, the ceiling function, and since at the end we take the limit for $m \rightarrow \infty$, the results remain unchanged.
\end{remark}

\noindent {\bf Phase~2}: This phase is similar to the previous one. The transmitter creates
\begin{align}
\frac{m}{1-\delta_{12}}
\end{align}
linearly independent equations of the $m$ bits for receiver $2$ and sends them out. For $(1-\delta_F)$ fraction of the time, feedback is available and
\begin{align}
(1-\delta_F) \frac{\delta-\delta_{12}}{1-\delta_{12}} m
\end{align}
bits are available at ${\sf Rx}_1$ and needed at ${\sf Rx}_2$. Denote these bits by $v_{2|1}$. Moreover, for $\delta_F$ fraction of the time, no feedback is available. But due to the statistics of the channel, we know
\begin{align}
\label{Eq:Rx2noFB}
\delta_F \frac{\delta-\delta_{12}}{1-\delta_{12}} m = \frac{\delta_F}{1-\delta_F} \left| v_{2|1} \right|
\end{align}
bits are available at ${\sf Rx}_1$ and needed at ${\sf Rx}_2$. The transmitter creates the same number of linearly independent equations as (\ref{Eq:Rx2noFB}) from the transmitted signal when feedback links were not available. Denote these equations by $v^{\mathrm{noFB}}_{2|1}$.

\noindent {\bf Phase~3}: In this phase, the transmitter encodes $v_{1|2}$ and $v_{2|1}$ using erasure codes of rate $(1-\delta)$. Note that $\left| v_{1|2} \right| = \left| v_{2|1} \right|$. The transmitter creates the XOR of the encoded bits for ${\sf Rx}_1$ and ${\sf Rx}_2$ and sends them out. This Phase takes
\begin{align}
\frac{\left| v_{1|2} \right|}{(1-\delta)} = \frac{(1-\delta_F)(\delta-\delta_{12})}{(1-\delta)(1-\delta_{12})}m.
\end{align}
time instants.

\noindent {\bf Recursive step}: Consider $v^{\mathrm{noFB}}_{1|2}$ and $v^{\mathrm{noFB}}_{2|1}$ as the input messages in new Phase~1 and Phase~2, respectively, and repeat the communication strategy.

\noindent {\bf Termination}: To simplify the protocol, when the remaining bits in $v^{\mathrm{noFB}}_{1|2}$ and $v^{\mathrm{noFB}}_{2|1}$ is $o\left( m^{1/3} \right)$ we stop the recursion and send the remaining bits using time sharing between the two erasure codes. Note that while we used $o\left( m^{1/3} \right)$ as our termination threshold, any threshold with vanishing (as $m \rightarrow \infty$) impact would work.

\noindent {\bf Decoding}: Our transmission protocol is built upon that of global feedback~\cite{GeorgiadisDetDelayedBC,Wang_12,GatzianasGeorgiadis_13} with the addition of the recursive step. Decoding starts with the last recursive step and goes backwards to the first iteration. A subtle point worth noting is that in each iteration the transmitter creates linearly independent equations similar to $v^{\mathrm{noFB}}_{1|2}$ and $v^{\mathrm{noFB}}_{2|1}$, and we need to guarantee that this task is feasible as we have many iterations. It is easy to verify that the geometric sum of the number of linearly independent equations created for each receiver during all iterations is smaller than the total number of unknown bits we start with, \emph{i.e.} $m$. Thus, the transmitter is able to carry out its task as needed.

\noindent {\bf Achievable rates}: Achievable rates are calculated as the ratio of the number of transmitted bits divided by the total communication time for $m \rightarrow \infty$ as below:
\begin{align}
\label{Eq:RatesCase2}
R_{\mathrm{SUM}} \overset{m \rightarrow \infty}= \frac{2m}{\frac{2 m}{1-\delta_{12}}+ \frac{(1-\delta_F)(\delta-\delta_{12})}{(1-\delta)(1-\delta_{12})} m + \frac{2 \left| \bar{v}^{\mathrm{noFB}}_{2|1} \right|}{R_{\mathrm{SUM}}}}.
\end{align}
where
\begin{align}
\label{Eq:RatesCase2Exp}
\left| \bar{v}^{\mathrm{noFB}}_{2|1} \right| = \delta_F \frac{\delta-\delta_{12}}{1-\delta_{12}} m.
\end{align}
From~\eqref{Eq:RatesCase2} and~\eqref{Eq:RatesCase2Exp}, we have
\begin{align}
& R_{\mathrm{SUM}} \left( \frac{2}{1-\delta_{12}}+ \frac{(1-\delta_F)(\delta-\delta_{12})}{(1-\delta)(1-\delta_{12})} \right) = 2 - 2 \delta_F \frac{\delta-\delta_{12}}{1-\delta_{12}} \nonumber \\
& \Rightarrow R_{\mathrm{SUM}} = \frac{2A(1-\delta)}{2(1-\delta)+(1-\delta_F)(\delta-\delta_{12})},
\end{align}
where for the last equality, we used the definition of $A$ as given in~\eqref{Def:A}:
\begin{align}
A = (1-\delta_{12}) - \delta_F(\delta-\delta_{12}).
\end{align}

We can further rearrange the denominator of this equality as
\begin{align}
2(1-\delta)+(1-\delta_F)(\delta-\delta_{12}) = (1-\delta) + (1-\delta_{12}) - \delta_F(\delta-\delta_{12}) \overset{\eqref{Def:A}}= 1 -\delta + A,
\end{align}
to obtain
\begin{align}
R_{\mathrm{SUM}} = R_1 + R_2 = \frac{2A(1-\delta)}{(1-\delta+A)},
\end{align}
where $R_1$ and $R_2$ are given in (\ref{Eq:TargetRates}). This completes the avhievability proof of Case~2, and thus, the proof of Theorem~\ref{THM:Ach_IFB}

%% file: Achievability_InnerBound_IFB.tex
In this section, we provide the proof of Theorem~\ref{THM:InnerBound_IFB} which establishes a connection to the Blind Index Coding problem~\cite{kao2016blind}.

\begin{figure}[!ht]
\centering
\subfigure[]{\includegraphics[width = .4\columnwidth]{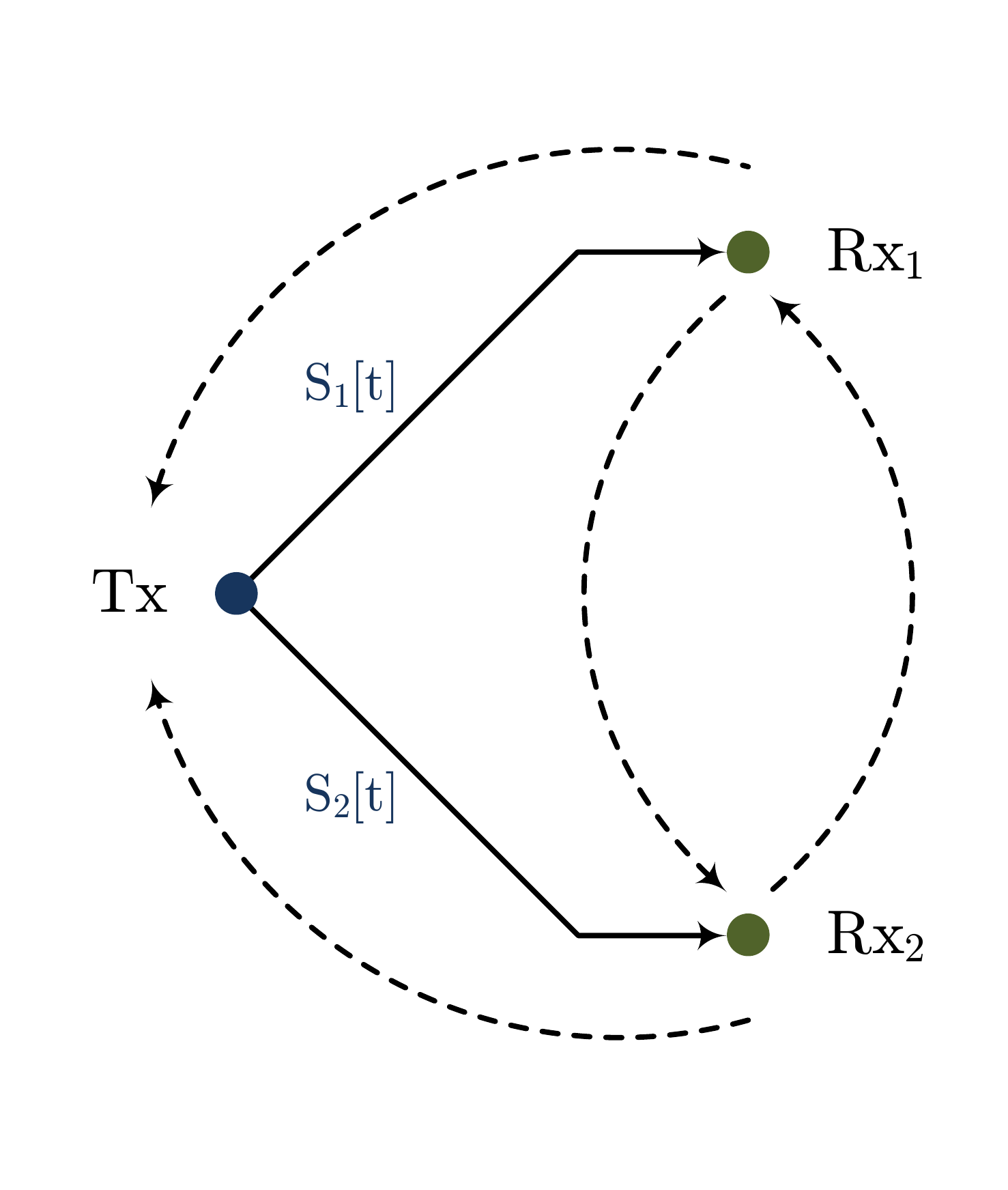}}
\hspace{.5in}
\subfigure[]{\includegraphics[width = .4\columnwidth]{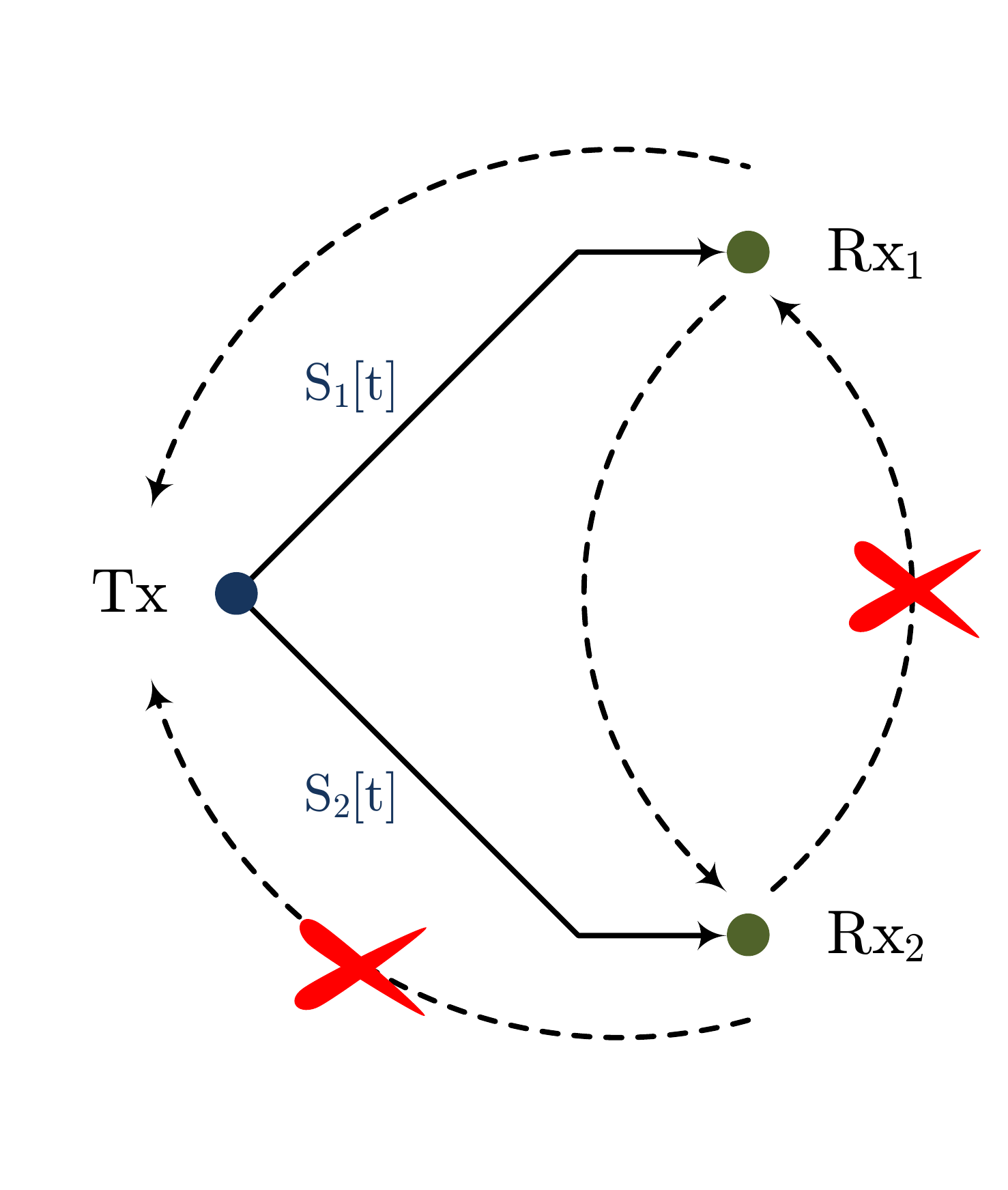}}
\subfigure[]{\includegraphics[width = .4\columnwidth]{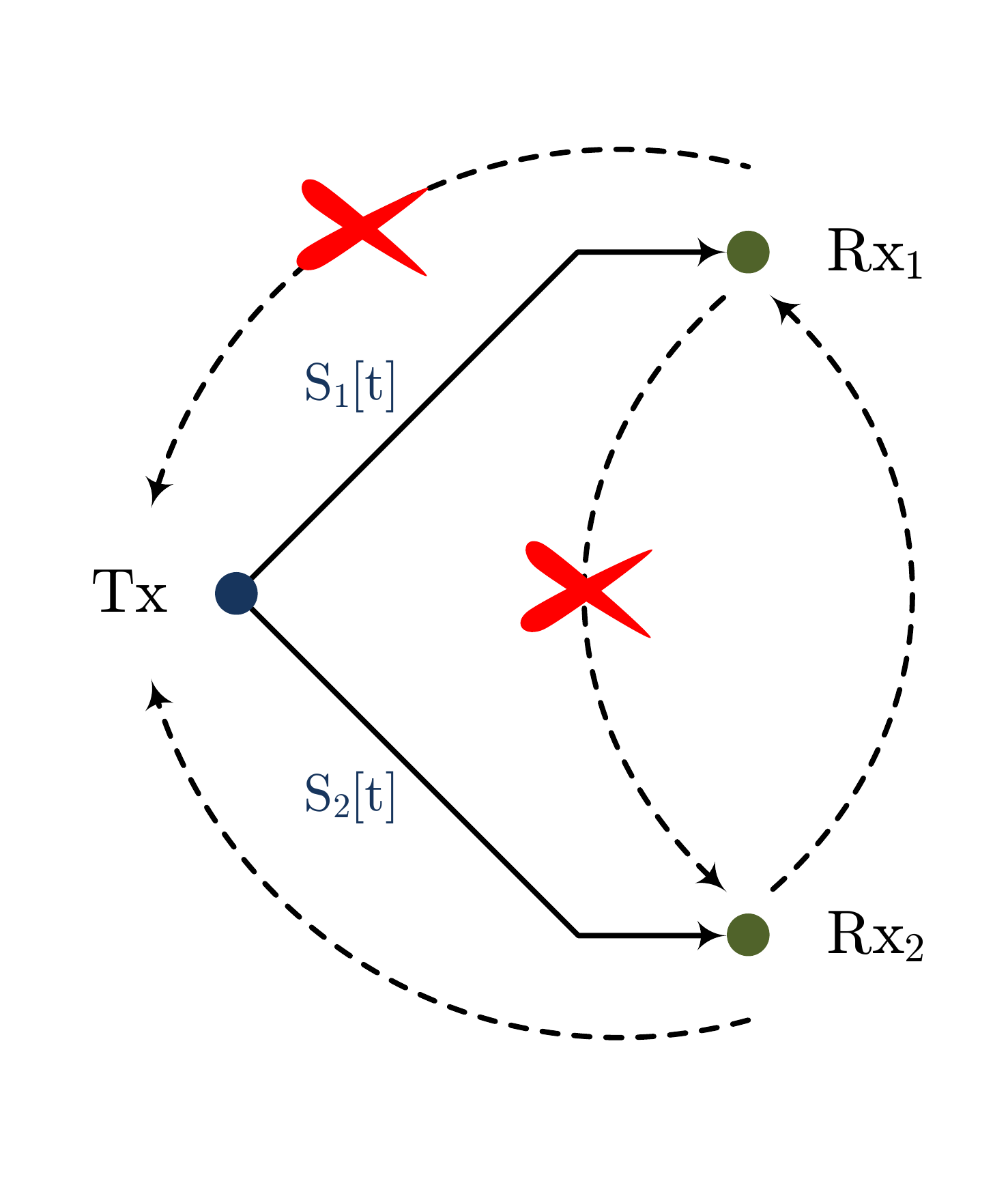}}
\hspace{.5in}
\subfigure[]{\includegraphics[width = .4\columnwidth]{FiguresPDF/BC-Intermittent-alloff.pdf}}
\caption{(a) All feedback links are available ($1-2\delta+\delta_{12}$); (b) feedback only from ${\sf Rx}_1$ ($\delta-\delta_{12}$); (c) feedback only from ${\sf Rx}_2$ ($\delta-\delta_{12}$); (d) all feedback signals are erased ($\delta_{12}$).}\label{Fig:FBCase3}
\end{figure}

Recall that in Case~2 of Theorem~\ref{THM:Ach_IFB}, feedback links are fully correlated, \emph{i.e.}
\begin{align}
\Pr\left( S_{F1T}[t] \neq S_{F2T}[t] \right) = 0.
\end{align}
In Theorem~\ref{THM:InnerBound_IFB}, on the other hand, we allow for a more general distribution of feedback links originating from different receivers. More specifically, feedback links from each receiver are still fully correlated, \emph{i.e.} $S_{F1T}[t] = S_{F12}[t]$ for $\msf{Rx}_1$ and vice versa, but feedback links originating from different receivers can have a more general distribution (unlike Case~2). In other words, feedback scenarios in Case~2 are always homogenous, while they can be heterogeneous now. Thus, we need new ideas inspired by those reviewed in the heterogeneous Case~1 of Theorem~\ref{THM:Ach_IFB} to design the transmission protocol. We also assume all erasure links are governed by Bernoulli $(1 - \delta)$ random variables and $\delta_{FF} = \delta_{12} = \delta^2$. Fig.~\ref{Fig:FBCase3} depicts all possible feedback realizations and the corresponding probabilities. The non-trivial corner point based on Theorem~\ref{THM:Capacity_IFB} is given by
\begin{align}
\label{Eq:RSUMCase3}
R_1 = R_2 = \frac{\beta(1-\delta)}{1+\beta},
\end{align}
where
\begin{align}
\label{Def:beta-case3}
\delta_i = \delta_{F_i} = \delta, \delta_{12} = \delta_{FF} = \delta^2 \Rightarrow \beta_i = \beta = \frac{\delta_{12}(1-\delta)+(1-\delta_{12})^2}{(1-\delta)} \overset{\delta < 1}= 1+\delta-\delta^3.
\end{align}
In this section, we show we can achieve
\begin{align}
\label{Eq:IFB_AchievableRates}
R_1 = R_2 = \frac{\left( 1 - \delta^2 \right)}{2+\delta+\delta^3}.
\end{align}

\noindent {\bf Specific assumptions of Theorem~\ref{THM:InnerBound_IFB}}: Before describing the protocol, we would like to highlight the specific channel setup in this case. First, we note that all channel links (forward and feedback) are Bernoulli $\left( 1 - \delta \right)$, and $S_{FiT}[t] = S_{Fi\bar{i}}[t]$, $i=1,2$. In particular, the latter assumption means that each receiver has access to (at least) what the transmitter knows about the channel state information. Finally, we assume $\delta_{FF} = \delta_{12} = \delta^2$. We note the transmission protocol in this case is highly tailored to these assumptions and generalization to other setting is not a straightforward task.


\noindent {\bf Transmission Protocol (overview)}: We start with $m$ bits for each user. The protocol includes an initial round that utilizes feedback to recycle bits and a BIC step which ignores feedback. The initial round includes two phases each dedicated to one user, and an XOR phase where bits that are available as side-information to unintended receivers are sent. The protocol is then followed by a BIC step where remaining bits are sent using erasure codes where we capitalize on the fact that statistically some bits for
${\sf Rx}_i$ are available at ${\sf Rx}_{\bar{i}}$, $i=1,2$. In Case~2 of Theorem~\ref{THM:Ach_IFB}, only two feedback realizations could happen captured in Fig.~\ref{Fig:FBCase3}(a) and Fig.~\ref{Fig:FBCase3}(d), and in the recursive step, we send random linear combination of the bits for which no feedback was received. In this case, we take advantage of the feedback structure and send individual equations that we know the intended receiver is {\it missing} and rely on the statistics of the channel to make some of these equations available to the other receiver.

\noindent {\bf Phase~1}: The transmitter creates
\begin{align}
\frac{m}{1-\delta_{12}}
\end{align}
linearly independent combinations of the $m$ bits for receiver $1$ and sends them out. We refer to these combinations as the coded bits for receiver $i$. When feedback channel realization $(a)$ of Fig.~\ref{Fig:FBCase3} occurs, which happens with frequency $\left( 1 - 2\delta + \delta_{12} \right)$,
\begin{align}
\label{Eq:Rx1FBa}
\underbrace{\left( 1 - 2\delta + \delta_{12} \right)}_{\text{realization a}} \frac{\delta-\delta_{12}}{1-\delta_{12}} m
\end{align}
coded bits are available at ${\sf Rx}_2$ and needed at ${\sf Rx}_1$. Denote these coded bits by $v^{(a)}_{1|2}$. When feedback channel realization $(c)$ of Fig.~\ref{Fig:FBCase3} occurs, which happens with frequency $\left( \delta - \delta_{12} \right)$, statistically
\begin{align}
\label{Eq:Rx1FBc}
\underbrace{\left( \delta - \delta_{12} \right)}_{\text{realization c}} \frac{\delta-\delta_{12}}{1-\delta_{12}} m
\end{align}
coded bits are available at ${\sf Rx}_2$ and needed at ${\sf Rx}_1$. Similar to the re-transmission of coded bits identified in Phase 2 of Case~1, the transmitter creates the same number random linear combinations of the coded bits that delivered to ${\sf Rx}_2$ under realization $c$ of Fig.~\ref{Fig:FBCase3}. Denote these random combinations by $v^{(c)}_{1|2}$. Note that both $v^{(a)}_{1|2}$ and $v^{(c)}_{1|2}$ are \underline{known to receiver $2$}  due to the available CSIR in \eqref{Eq:DecodingFunction} (recalling $S_{F12}[t]=S_{F1T}[t]$ in our setting), and further the transmitter and receivers can share the matrices that are used to create random combination prior to the beginning of the communication block. Since $v^{(a)}_{1|2}$ and $v^{(c)}_{1|2}$ are similar in nature, for simplicity, we combine/concatenate them into $v_{1|2}$ of size
\begin{align}
\label{Eq:Rx1FB}
\underbrace{(1-\delta)}_{\text{realizations a $\&$ c}} \frac{\delta-\delta_{12}}{1-\delta_{12}} m
\end{align}

Moreover, during the Phase 1, for $\delta - \delta_{12}$ fraction of the time, feedback realization $(b)$ of Fig.~\ref{Fig:FBCase3} occurs and due to the statistics of the channel, we know
\begin{align}
\label{Eq:Rx1noFB}
\delta\frac{\delta-\delta_{12}}{1-\delta_{12}} m = \frac{\delta}{1-\delta} \left| v_{1|2} \right|
\end{align}
coded bits are {\it missing} at ${\sf Rx}_1$. Note that a fraction $\left( 1- \delta \right)$ of these combinations is available at ${\sf Rx}_2$. The transmitter selects the coded bits missing at ${\sf Rx}_1$ as Phase 1 of Case~1 when feedback realization $(b)$ of Fig.~\ref{Fig:FBCase3} occurs, and denote them by $v^{\mathrm{noFB}}_{1|2}$. Thus, we have
\begin{align}
\left| v^{\mathrm{noFB}}_{1|2} \right| = \delta\frac{\delta-\delta_{12}}{1-\delta_{12}} m = \frac{\delta}{1-\delta} \left| v_{1|2} \right|.
\end{align}

\begin{remark}
Here, we would like to point out a subtle but important distinction between combinations in $v_{1|2}$ and in $v^{\mathrm{noFB}}_{1|2}$. In the former, we effectively create random combinations of previously transmitted coded bits in Phase~1 that are known to ${\sf Rx}_2$. In the latter, for $v^{\mathrm{noFB}}_{1|2}$, the transmitter selects missing individual coded bits in Phase~1  according to the feedback, and a fraction $\left( 1- \delta \right)$ of these are already known to ${\sf Rx}_2$. We will use the structure of $v^{\mathrm{noFB}}_{1|2}$ to reduce the communication length of the initial phase in the following iterations as discussed below.
\end{remark}

\noindent {\bf Phase~2}: The transmission is same as that in Phase~1 by swapping user index 1 with index 2, and the transmitter creates $v_{2|1}$ and $v^{\mathrm{noFB}}_{2|1}$ for re-transmission.

\noindent {\bf XOR~Phase}: In this phase, the transmitter encodes $v_{1|2}$ and $v_{2|1}$ using erasure codes of rate $(1-\delta)$. Note that $\left| v_{1|2} \right| = \left| v_{2|1} \right|$. The transmitter creates the XOR of the encoded bits for ${\sf Rx}_1$ and ${\sf Rx}_2$ and sends them out. This Phase takes
\begin{align}
\frac{\left| v_{1|2} \right|}{(1-\delta)} = \frac{(1-\delta_F)(\delta-\delta_{12})}{(1-\delta)(1-\delta_{12})}m \overset{\delta_F = \delta}= \frac{(\delta-\delta_{12})}{(1-\delta_{12})}m.
\end{align}
time instants.

\noindent {\bf Blind Index Coding Phase}: The main difference between the achievability strategies of Case~2 of Theorem~\ref{THM:Ach_IFB} and this case lies in this re-transmission Phase, and the protocol is no longer recursive. In this case, $v^{\mathrm{noFB}}_{1|2}$ and $v^{\mathrm{noFB}}_{2|1}$, unlike in Case~2, are not random linear combinations of previously transmitted coded bits, but rather are individual coded bits that we know the intended receiver is missing and the unintended receiver knows a fraction of them. We note that $|v^{\mathrm{noFB}}_{1|2}| = |v^{\mathrm{noFB}}_{2|1}|$ is given in \eqref{Eq:Rx1noFB}, and due to the specific choice of channel parameters of Theorem~\ref{THM:InnerBound_IFB}, we do not need to recycle bits from realization $(d)$ of Fig.~\ref{Fig:FBCase3}.  Further, due to the statistics of the channel, a fraction $(1-\delta)$ of these missing coded bits are already available to the unintended receiver. Thus, this step falls under the broad definition of the Blind Index Coding which we will further discuss in the next section.

The available side information means that if the transmitter provides each receiver a total of
\begin{align}
\left( 2-(1 - \delta) \right) \times \underbrace{\delta \frac{\delta-\delta_{12}}{1-\delta_{12}} m}_{=~|v^{\mathrm{noFB}}_{1|2}|}
\end{align}
random linear combination of $v^{\mathrm{noFB}}_{1|2}$ and $v^{\mathrm{noFB}}_{2|1}$, then, each receiver will have enough equations to recover \underline{both} $v^{\mathrm{noFB}}_{1|2}$ and $v^{\mathrm{noFB}}_{2|1}$. The transmitter creates these random combinations and encodes them at an erasure code of rate $\left( 1 - \delta \right)$ and multicast them out. We note that the transmitter does not utilize the feedback in this step. In the next section, we discuss why this decision results in sub-optimal rates. This step takes
\begin{align}
\frac{\delta \left( 1 + \delta \right) \left( \delta-\delta_{12} \right)}{\left( 1 - \delta \right) \left(1-\delta_{12} \right)} m
\end{align}
time instants, and at the end of it, $v^{\mathrm{noFB}}_{1|2}$ and $v^{\mathrm{noFB}}_{2|1}$ becomes available to both receivers.

\noindent {\bf Decoding}: We use receiver 1 as an example. After the BIC phase, ${\sf Rx}_1$ decodes both $v^{\mathrm{noFB}}_{1|2}$ and $v^{\mathrm{noFB}}_{2|1}$. Then, using the available $v^{\mathrm{noFB}}_{1|\bar{2}}$ it completes the recovery of all 
\begin{align}
\label{Eq:DecodingCaseb}
\frac{\delta-\delta_{12}}{1-\delta_{12}}m
\end{align}
coded bits intended for ${\sf Rx}_1$ and transmitted in Case (b) of Figure~\ref{Fig:FBCase3}. Next, using $v^{(a)}_{1|2}$ obtained from the XOR phase, ${\sf Rx}_1$ completes the recovery of
\begin{align}
\label{Eq:DecodingCasea}
(1-2\delta+\delta_{12})\frac{1-\delta_{12}}{1-\delta_{12}}m
\end{align} 
coded bits transmitted in Case (a) of Figure~\ref{Fig:FBCase3}; also, using linear combinations $v^{(c)}_{1|2}$, ${\sf Rx}_1$ can decode all coded bits when ${\sf Rx}_2$ is on in Case (c) of Figure~\ref{Fig:FBCase3}, thus obtaining  
\begin{align}
\label{Eq:DecodingCasec}
(\delta-\delta_{12})\frac{1-\delta_{12}}{1-\delta_{12}}m
\end{align} 
code bits in this case. Finally, ${\sf Rx}_1$ receives 
\begin{align}
\label{Eq:DecodingCased}
(1-\delta)\frac{\delta_{12}}{1-\delta_{12}}m
\end{align}
coded bits in Case (d) of Figure~\ref{Fig:FBCase3}. Adding up \eqref{Eq:DecodingCasea}--\eqref{Eq:DecodingCased},  ${\sf Rx}_1$ gathers a total of $m$ coded bits, and thus, the decodability at $\msf{Rx}1$ is ensured.


\noindent {\bf Achievable rates}:
The total communication time (ignoring the approximate terms discussed in Remark~\ref{remark:Bernstein}) can be then calculated as
\begin{align}
T = \underbrace{\frac{2m}{(1-\delta_{12})}}_{\mathrm{ph.~1~and~ph.~2}} + \underbrace{\frac{(\delta-\delta_{12})m}{(1-\delta_{12})}}_{\mathrm{XOR~ph}} + \underbrace{\frac{\delta \left( 1 + \delta \right) \left( \delta-\delta_{12} \right)}{\left( 1 - \delta \right) \left(1-\delta_{12} \right)} m}_{\mathrm{BIC~ph}} = \frac{2+\delta+\delta^3}{\left( 1 - \delta^2 \right)}m.
\end{align}
The achievable rates are then calculated as $R_i = m/T$ which coincide with the expression given in (\ref{Eq:IFB_AchievableRates}). This completes the proof of Theorem~\ref{THM:InnerBound_IFB}. We compared these achievable rates to the outer-bounds of Theorem~\ref{THM:Capacity_IFB} in Section~\ref{Section:Main_IFB}, Figure~\ref{Fig:InnerOuterIFB}.